\documentclass[5p,11pt]{elsarticle}

\usepackage[english]{babel}
\usepackage[T1]{fontenc}
\usepackage{amsfonts}
\usepackage[normalem]{ulem}
\usepackage{amsmath}
\usepackage{amssymb}
\usepackage{amsbsy}
\usepackage{xcolor}
\usepackage{hyperref}
\usepackage{graphicx}
\usepackage{setspace}
\usepackage{xspace}
\usepackage{array}
\usepackage{lipsum}
 \usepackage{algorithm}
 \usepackage{algorithmicx}
 \usepackage{algpseudocode}
\usepackage{bbold}
\usepackage{tabularx}
\usepackage{multirow}
\usepackage[group-separator={,},round-mode=places,round-precision=1,group-minimum-digits=4]{siunitx}
\usepackage{booktabs} 
\newcolumntype{L}{>{$}l<{$}}
\newcolumntype{C}{>{$}c<{$}}
\newcolumntype{R}{>{$}r<{$}}

\usepackage{pgf}
\usepackage{tikz}
\usetikzlibrary{shapes,snakes,arrows,patterns,matrix,positioning}
\usepackage{pgfplots}
\usepackage{pgfplotstable}
\usepackage{subfig}

\begin{document}

%
%
\title{Solving the Bethe-Salpeter equation on massively parallel architectures}

\author[ads0]{Xiao Zhang\corref{cor2}}
\ead{xzhng125@illinois.edu}

\author[ads1,ads2]{Sebastian Achilles}
\ead{s.achilles@fz-juelich.de}

\author[ads2]{Jan Winkelmann}
\ead{winkelmann@aices.rwth-aachen.de}

\author[ads3]{Roland Haas}
\ead{rhaas@ncsa.illinois.edu}

\author[ads3,ads4,ads5]{Andr\'e Schleife\corref{cor2}}
\ead{schleife@illinois.edu}

\author[ads1]{Edoardo Di Napoli\corref{cor1}}
\ead{e.di.napoli@fz-juelich.de}

\address[ads0]{Department of Mechanical Science and Engineering, University of Illinois at Urbana-Champaign, Urbana, IL 61801, USA.}
\address[ads1]{J\"ulich Supercomputing Centre, Forschungszentrum J\"ulich, Wilhelm-Johnen-Stra\ss e, 52425 J\"ulich, Germany.}  
\address[ads2]{RWTH Aachen University, Aachen Institute for Advanced Study in Computational Engineering Science, Schinkelstr.\ 2, 52062 Aachen, Germany.}
\address[ads3]{National Center for Supercomputing Applications, University of Illinois at Urbana-Champaign, Urbana, IL 61801, USA.}
\address[ads4]{Department of Materials Science and Engineering, University of Illinois at Urbana-Champaign, Urbana, IL 61801, USA.}
\address[ads5]{Materials Research Laboratory, University of Illinois at Urbana-Champaign, Urbana, IL 61801, USA.}

\cortext[cor1]{Principal corresponding author}
\cortext[cor2]{Corresponding author}






\begin{abstract}
  The last ten years have witnessed fast spreading of massively parallel computing clusters, from leading supercomputing facilities down to the average university computing center.
  Many companies in the private sector have undergone a similar evolution.
  In this scenario, the seamless integration of software and middleware libraries is a key ingredient to ensure portability of scientific codes and guarantees them an extended lifetime.
  In this work, we describe the integration of the ChASE library, a modern parallel eigensolver, into an existing legacy code for the first-principles computation of optical properties of materials via solution of the Bethe-Salpeter equation for the optical polarization function.
  Our numerical tests show that, as a result of integrating ChASE and parallelizing the reading routine, the code experiences a remarkable speedup and greatly improved scaling behavior on both multi- and many-core architectures.
  We demonstrate that such a modernized BSE code will, by fully exploiting parallel computing architectures and file systems, enable domain scientists to accurately study complex material systems that were not accessible before.
\end{abstract}


\maketitle

\pagenumbering{arabic}
\renewcommand{\thefootnote}{\arabic{footnote}}

%
%




\section{Introduction}

As the size of massively parallel computing clusters is advancing
towards the exascale regime, utilizing the great power of such
clusters for scientific applications is a very important task.  In the
last decade, modernizing domain-specific software by improving its
parallelism and efficiency has grown into a mainstream activity in
many fields of computational scientific research.  Co-design and
portability is playing an ever increasing role in coordinating the
effort in pushing the development of new hardware, efficient
middleware, such as numerical libraries, and their use to run
massively parallelized simulations~\cite{anzt2020, bernholdt2020,
  kothe2018, Shao:2018}.  In this paper, we present a notable example
of such an effort. We describe how the initialization of a large and
dense Hermitian eigenproblem and the solution for a small portion of
its spectrum are the two major bottlenecks in the parallel execution
of an existing code \cite{fuchs08,Roedl:2008} for solving
the Bethe-Salpeter equation (BSE) \cite{salpeter1951, onida02}.
 We illustrate how we
improve the parallel I/O and integrate a modern highly parallelized
solver---the Chebyshev Accelerated Subspace Iteration Eigensolver
(ChASE)~\cite{winkelmann2019chase}---into the solution of the BSE.
Besides increased performance and parallelism, we demonstrate that our
implementation allows the code to tackle physical problems that could
not be addressed before. 

In the context of this work, the solution of the BSE equation is used
to study the response of a material to an external electromagnetic
field---including visible light---and in so doing derive the material
optical properties.  Accurate and predictive modeling of optical
response is important, for instance, to understand the functionality
of optoelectronic devices, whose applications are directly related to
light absorption, reflection, and transmission.  For instance,
creation or separation of excitons is an important process in
photovoltaic cells and light-emitting diodes
\cite{miyata2015direct,klingshirn2007zno}. Upon optical excitation via
an electromagnetic field, electrons in a material leave their original
electronic ground state, gain energy, and reach an excited state.
This process leaves a positively charged hole in the initially
occupied valence state.  The electrons and holes couple via the
Coulomb interaction as they propagate, rendering this situation an
intricate electron-hole quantum mechanical problem that is much more
complicated than the ground state.

While these optical
properties can be measured experimentally by approaches including
ellipsometry \cite{drude1887ueber},
photoluminescence~\cite{perkowitz2012optical}, and
photoemission~\cite{hufner2013photoelectron}, accurate predictive
studies are crucial in interpreting experiments and guiding research
towards successful discovery of materials for different applications.
Such predictive \emph{in silico} studies of optical properties require
a description of excited electronic states and are carried out using
first-principles approaches that solve approximations to the
fundamental Schr\"{o}dinger equation in a non-parametric fashion.
This would then allow one to carry out quantitative simulations to
compute material properties. The BSE approach is
based on many-body perturbation theory (MBPT) and is a common
first-principles method to simulate the underlying optical absorption
strength and its energy dependence accurately.

This Green's function based approach considers the external electric
field as a perturbation to the electronic ground state and describes
the response of the electronic system to such a perturbation.  Key to
the effectiveness of the approach is that it accounts for the
\emph{screened} Coulomb interaction between the electron and the hole,
and thus gives an accurate account of the optical excitation process.
However, one of the main challenges of this approach lies in its
extremely large computational cost which scales as
$\mathcal{O}(n_e^6)$~\cite{Ljungberg2015}, where $n_e$ is the number of
electrons in the material.  Although this can be reduced to
$\mathcal{O}(n_e^3)$ or even $\mathcal{O}(n_e^2)$ with advanced
solvers~\cite{Ljungberg2015, fuchs08}, studying large or complicated
material systems using the BSE approach is still limited by the large
size of the corresponding eigenproblem, which can go up to
$\mathcal{O}(10^6)$.

In the last decade, tier-0 computing clusters have grown to the point
of having hundreds of thousands of computing cores with peak
performances up to hundreds of Peta-FLOPS (Floating Point Operations
per Second).  Such massively parallel
architectures have the potential to tackle and solve eigenproblems
significantly larger than $\mathcal{O}(10^6)$. Therefore, solving for
the largest BSE is quite within the capability of the available
hardware. The real challenge is the ability of numerical libraries to
take advantage of such computing resources and parallelize
efficiently over many thousands of cores for a given eigenproblem.

One of the disadvantages of dense eigenvalue solvers is that their
complexity usually scales as the cubic power of the matrix size $N$,
while their memory footprint grows as $N^2$.  This limits the rapid
scaling of the problem size to a large number of computing cores.  On
the other hand, in many important cases the BSE eigenproblem has to be
solved for a very small fraction of its spectrum, usually comparable
to or less than 1\,\% of its total size.

Numerical libraries such as ScaLAPACK~\cite{slug} or the more modern
SLATE~\cite{slate} maintain a very high complexity even in those cases
where the used algorithm can compute a portion of the spectrum---e.g.\
the MRRR eigensolver. Conversely, iterative eigensolvers, such as
Conjugate Gradient (CG)~\cite{Geradin, Fried} or
Lanczos~\cite{Cullum:1985wj}, could in principle have a lower
complexity closer to $N^2$.  When implemented in iterative libraries
such as SLEPc~\cite{Hernandez:2005:SSF} or
Trillinos~\cite{Trillinos:1089021}, these iterative algorithms pay the
price of a smaller arithmetic intensity. Few operation per byte are
executed with the result that the processor remains idle waiting for
the necessary data to be moved through the memory hierarchy. The
resulting number of operations per second is rather low with respect
to the peak performance of the CPU. It is customary to refer to the
executed operations as ``slow''
FLOPs\footnote{To be distinguished from FLOPS which is a rate of
  operations per second}.

In this paper we argue that there are alternatives to both high
complexity and ``slow'' FLOPs. We show how a modern library based on
subspace iteration and augmented with Chebyshev filter can be a
winning alternative in cases where a limited portion of the spectrum
of a large dense eigenvalue problem is desired.  The ChASE library has
a complexity of $\mathcal{O}(mN^2)$, with $m$ proportional to the
number of eigenpairs sought after.  When $m \ll N$, ChASE executes a
number of operations to compute the solution that is substantially smaller
than $\mathcal{O}(N^3)$.  At the same time, most of the ChASE floating
point operations ($\sim 90\%$) are executed using BLAS level 3
subroutines that are known to extract most of the peak performance out
of any existing parallel computing architecture.

We integrated the highly parallelized ChASE library with an existing
BSE code that has
its origin in the group of Friedhelm Bechstedt at the
Friedrich-Schiller University Jena, where it was developed over many
years \cite{Hahn:2001,Schmidt:2003}. It is closely developed around the Vienna Ab-Initio Simulation Package (VASP) and its implementation
is described in Refs.~\cite{fuchs08,Roedl:2008}. The integration of ChASE leads
to significant improvement when the BSE code is executed on massively
parallel clusters. On a given set of computing resources, the gain in
adopting ChASE can exceed a factor of 4\,--\,5 in runtime compared to
the previously used CG solver. 
Comparing the parallel efficiency of both solvers supports such
result: Chase's efficiency curve maintains much higher values than CG
as the number of computing nodes increases. In addition,
the higher performance outcome of ChASE allows us to explore material
systems where the previous implementation could not fully unravel the
sought after physics. We illustrate this feature by looking into the
extraction of accurate optical properties of the organic crystal
naphthalene.

The paper is organized as follows. In
Sec.~\ref{sec:physics}, we explain the details of the BSE approach and
describe the challenges of dealing with the BSE problem due to the
necessity of solving large size matrices.  In Sec.~\ref{sec:compute},
we illustrate how such challenges appear in our existing BSE code,
both in the I/O procedure and the eigensolver and explain the many
benefits of the new I/O routine and the ChASE solver.
In Sec.~\ref{sec:results}, we describe numerical experiments to
measure the strong and weak scaling and illustrate the power of
ChASE in addressing optoelectronic properties of complex
materials. Section \ref{sec:conclusion} summarizes and concludes the
paper.


\section{\label{sec:physics}Modeling the Optical Properties of Materials}
Solving the Bethe-Salpeter equation (BSE) \cite{salpeter1951} for the optical polarization function defines a theoretical spectroscopy approach that accounts for two-particle, electron-hole excitations, by including the interaction between excited electrons and holes using many-body perturbation theory \cite{onida02}.
It is a first-principles framework used to accurately predict fundamental optical properties, such as optical absorption spectra including excitonic effects and exciton binding energies for molecules, semiconductors, and insulators.
The exciton binding energy is defined as the difference between interacting and non-interacting electron-hole pair energies and corresponds to the energy required to separate a bound electron-hole pair.

The BSE is based on a Green's function technique to describe the electron-electron interaction and can be derived from Hedin's system of equations \cite{Hedin:1965} via an expansion of the electronic self energy $\Sigma$ into increasing orders of the screened electron-electron interaction $W$ \cite{Louie:2006,Bechstedt:2015}.
For numerical implementations the solution of the BSE is transformed into an eigenvalue problem $\hat{H}^{\text{BSE}}|\Lambda\rangle=E_\Lambda|\Lambda\rangle$, with the exciton Hamiltonian~\cite{onida02,Bechstedt:2015,fuchs08}
\begin{equation}
\label{bseH}
\hat{H}^{\text{BSE}}_{vc\mathbf{k},v'c'\mathbf{k'}}=(E_{c\mathbf{k}}-E_{v\mathbf{k}})\delta_{vv'}\delta_{cc'}\delta_{\mathbf{kk'}}+2\bar{v}^{v'c'\mathbf{k}'}_{vc\mathbf{k}}-W^{v'c'\mathbf{k}'}_{vc\mathbf{k}}.
\end{equation}
Here, the $c$, $v$, and $\mathbf{k}$ indices label conduction bands, valence bands, and points in reciprocal space, respectively.
$E_{c\mathbf{k}}$ and $E_{v\mathbf{k}}$ are energies of single-(quasi)particle electron and hole states and are computed by applying quasiparticle corrections to Kohn-Sham eigenvalues from ground-state density-functional theory \cite{Hohenberg:1964fz,Kohn:1965}.
It is common practice to compute these using one of two approaches:
(i) The single-quasiparticle Green's function scheme within many-body perturbation theory, known as the $GW$ approximation of the electronic self energy \cite{Hedin:1965,onida02}, or (ii) hybrid exchange-correlation functionals in a generalized Kohn-Sham scheme \cite{Seidl:1996}.
Finally, the unscreened Coulomb interaction, denoted by $\bar{v}$ in Eq.\ \eqref{bseH}, represents short-range electron-hole exchange to account for local-field effects.
$W$ denotes the screened Coulomb interaction and requires an approximation and numerical description of dielectric screening. 

Solving the BSE numerically is computationally intensive:
The $\bar{v}$ and $W$ terms involve coupling between all electron-hole pairs and, as can be seen from Eq.~\eqref{bseH}, the rank of the BSE Hamiltonian depends on the number of (occupied) valence bands, (unoccupied) conduction bands, and the $\mathbf{k}$-point sampling of reciprocal space.
Of these, the number of occupied bands is determined by the number of electrons in the simulation cell for a given material.
However, the number of conduction bands and sampling of reciprocal space are convergence parameters.
While finer reciprocal-space sampling and more conduction bands lead to better converged solutions of the BSE, this increases the size of the BSE Hamiltonian.
In practice, one examines the convergence of physical quantities, such as the exciton-binding energy, with respect to a finite-sized sampling of reciprocal space.
To limit the number of empty states taken into account for the Hamiltonian, a BSE energy cutoff $E_{\text{cut}}$ is introduced such that only electron-hole pairs with $E_{c\mathbf{k}}-E_{v\mathbf{k}}<E_{\text{cut}}$ contribute.
Effects from electron-hole pairs with higher energies are typically small due to decreasing coupling $\bar{v}$ and $W$ and, therefore, can be neglected.
However, it needs to be tested what value for $E_{\text{cut}}$ achieves a certain convergence level, e.g.\ for the exciton-binding energy.

Even when $\mathbf{k}$-point sampling and BSE energy cutoff are
carefully converged, the resulting eigenproblem can become very large.
Its size increases linearly with the number of conduction bands, valence bands, and $\mathbf{k}$-points, as more non-interacting conduction-valence band pairs are included.
Depending on the exact goals of a simulation, e.g.\ whether just exciton binding energies or an entire spectrum are of interest, the typical test range for the BSE cutoff is on the order of 5\,--\,20 eV. 
The precise increase of the rank of the BSE matrix depends on details
of the electronic band structure of a material, and it typically increases a few orders of magnitude across this energy range, from several thousands to close to half a million.
For instance, in previous work we demonstrated that solving for eigenproblems with size as large as \num{360000} is necessary to obtain reasonable results for the convergence with respect to $\mathbf{k}$-point sampling \cite{fuchs08,schleife2009optical,schleife2018optical,zhang18}.
  
As an illustration, we describe below our previous work on optical properties of a meta-stable ZnO polymorph \cite{zhang18}.
To converge exciton binding energies with respect to $\mathbf{k}$-point sampling, we follow the procedure of Ref.\ \cite{fuchs08}:
For a number of samplings, the calculated binding energy is plotted as a function of the inverse number of $\mathbf{k}$-points.
The convergence in such a plot
shows linear behavior for very dense $\mathbf{k}$-point samplings, which allows an extrapolation to estimate the exciton-binding energy at infinitely dense sampling.
For ZnO, this linear regime is not observed until the inverse number
of $\mathbf{k}$-points is smaller than 0.015, corresponding to 66.7
$\mathbf{k}$-points in one specific reciprocal-space direction (see details in Ref.\ \cite{zhang18}).
The smallest inverse number of $\mathbf{k}$-points that we studied for this material in Ref.\ \cite{zhang18}, i.e., the densest $\mathbf{k}$-point sampling, is 0.012, corresponding to 83.3 $\mathbf{k}$-points in that direction.
This leads to
an eigenvalue problem with a size of $\sim\num{200000}$
and reading such a matrix and solving for 100 eigenvalues using the KSCG solver required 5\,--\,6 hours using 32 nodes of the BlueWaters supercomputer.
Pushing towards better convergence with respect to $\mathbf{k}$-point sampling
becomes expensive quickly:
Reducing the inverse number of $\mathbf{k}$-points to 0.010 increases the matrix size to $\sim\num{300000}$ and requires more than twice the memory.

The increase in matrix size is even more dramatic for more complex materials.
While the primitive unit cell of ZnO contains only four atoms and 36 valence electrons, these numbers can easily be one order of magnitude larger.
For instance, the unit cell of In$_2$O$_3$~\cite{schleife2018optical} contains 16 In and 24 O atoms, leading to 352 valence electrons.
In this case the matrix rank already amounts to $\sim\num{360000}$ using a $5\times5\times5$ $\mathbf{k}$-point grid and $E_{\text{cut}}=12.5$ eV.
Hence, computing converged optical spectra constitutes a serious computational problem, since it requires a converged $\mathbf{k}$-point grid \emph{and} a large BSE energy cutoff.
In this case, the BSE cutoff is determined by the maximum energy in the spectrum, which is typically much larger than what is needed to merely converge excitonic effects.
We mitigate this issue by using dense $\mathbf{k}$-point grids for the low energy range where a smaller BSE cutoff is sufficient, and more coarse $\mathbf{k}$-point grids for large photon energies, which requires large BSE cutoffs (see e.g.\ Ref.~\cite{schleife2018optical} for optical spectra of In$_2$O$_3$).
However, for materials with large unit cells, such as the organic crystal naphthalene C$_{20}$H$_{16}$ (see Sec.~\ref{sec:phys_problem}), computing exciton binding energies leads to large matrices on the order of $5\times10^5$ when converging $E_{\text{cut}}$ and $\mathbf{k}$-point sampling.

When computing exciton binding energies of a material, only a very small portion of the lower eigenspectrum is needed.
This is because studies of excitonic properties typically focus on states at or near the absorption edge, corresponding to the lowest eigenvalues of the BSE matrix.
When only few extremal eigenpairs are computed, it is customary to use iterative solvers even when the matrix
defining the eigenproblem is dense.
Such a choice is in part influenced by the overall complexity of the
iterative algorithm compared with a so-called direct one (e.g.\
Multiple Relatively Robust Representations~\cite{Dhillon:2004hx,
  Bientinesi:2005fq}).  In an iterative solver the total number of
floating point operations to reach the solution, is determined by the
complexity of the algorithm per iteration multiplied by the number of
iterations needed to converge. Since the overall number of iterations
is unknown a priori, the choice of iterative vs.\ direct solver
depends on the properties of the eigenproblem, the parameters of the
solver and its parallel efficiency, and is often a question of
practice and experience.

In our BSE code we relied on an iterative solver based on the Kalkreuther-Simma Conjugate-Gradient (KSCG) algorithm\cite{fuchs08,kalkreuter1996accelerated} to solve increasingly larger BSE eigenproblems.
While this algorithm presents clear advantages with respect to any direct solver, it lags behind when it is employed over increasingly larger parallel platforms. 
In the next section, we dig into the reasons for such lack of parallel performance and propose a modern alternative which is more  efficient, scales over massively parallel architectures, and enables us to tackle eigenproblems of unprecedented size.


\section{\label{sec:compute}The Computational Challenges}

The overall workflow towards the solution of a single BSE starts with
a density functional theory calculation to obtain single-particle
Kohn-Sham states and dipole matrix elements.  We perform these steps
using the Vienna Ab initio Simulation Package (VASP)
\cite{kresse1993,kresse19961,kresse19962}.  The VASP code is a
commercial open-source code commonly used for first-principles
calculations, and provides reliable results to compute the electronic
structures of materials.  The Kohn-Sham electronic structure and quasiparticle corrections are
then used to compute the BSE Hamiltonian, Eq.\
(\ref{bseH}).  Subsequently, either the lowest eigenvalues of the BSE
matrix are extracted by using an eigensolver based on the Conjugate
Gradient algorithm \cite{kalkreuter1996accelerated} or optical spectra
are computed using the time-propagation technique described in Refs.\
\cite{Hahn:2001,Schmidt:2003}.

\subsection{A brief introduction to the BSE code}

To obtain the solution to the BSE eigenvalue problem, there are two main stages that necessitate a large amount of computational resources:
the initialization of the BSE Hamiltonian and the computation of the lowest portion of its spectrum.
In our BSE implementation, described in detail in Refs.\ \cite{Roedl:2008,fuchs08}, the initialization is split into two steps:
1) the generation of the matrix elements as they are represented in Eq.\ (\ref{bseH}) and
the
assembly of the matrix in main memory as input for the solver.
The workflow for the first step of the initialization is embarrassingly parallel since there is no communication between different threads.
Thanks to the fact that the matrix elements in the BSE eigenvalue problem are independent of each other, the matrix elements are evaluated in chunks, where every subset of the matrix is computed independently from the others.
The user specifies the total number of jobs used to write the full matrix.
The calculation of the matrix elements is split between these multiple independent jobs, each parallelized over one entire node using OpenMP and writing a portion of the matrix on individual binary files.
In the second step, after all matrix elements are calculated and stored, a separate procedure reads in the matrix elements and performs the diagonalization (to compute eigenvalues) or a time-propagation scheme (to compute the dielectric function).
This scheme is beneficial if multiple different diagonalizations are to be performed on the same matrix, e.g.\ for convergence tests as described in Ref.\ \cite{fuchs08}.
Furthermore, as long as the time spent writing and reading the matrix is small compared to initialization and diagonalization steps, this scheme also allows efficient writing of large matrices using many single-node jobs and benefiting from backfill algorithms of modern queuing systems.
The workflow above is implemented in our BSE code \cite{fuchs08,Roedl:2008} discussed here.

As an illustration,
calculating the matrix elements for the ZnO system \cite{zhang18} for
a matrix of size \num{199433} requires about 600 node hours on Blue Waters,
while reading the matrix and computing the lowest 100 eigenvalues requires 39 node hours,
and 48 node hours (i.e., 87.3\%, 5.7\%, and 7.0\%), respectively.
The step of computing the matrix elements can be trivially parallelized on as many nodes as necessary.
However, the rest of the computational time is roughly split in half between I/O and solution of the eigenproblem.
Furthermore, in some of our tests the I/O alone required close to the maximum walltime on BlueWaters, making it challenging to complete these runs.
Consequently, the main challenges in improving the code lays in increasing the performance and parallelism of these two tasks.
In the following part of this section, we introduce the new parallel
reading process of the matrix elements to replace the old sequential
reading process, and a new sub-space iterative solver to improve the
parallel efficiency in computing the solution with respect to the conjugate gradient solver.

\subsection{Matrix generation and the I/O challenge}
\label{sec:IOdistr}
As discussed above, in step one each independent thread computes a number of rows of the exciton Hamiltonian matrix, Eq.\ (\ref{bseH}), and writes these to a single file.
Next, these individual files are read and the data is distributed amongst several MPI ranks for the subsequent solver step.
The existing implementation serialized reading data from files by assigning a ``reader'' rank to each file and allowing only one reader to read at any given time.
After completing reading its assigned file, the reader would broadcast the read data to all ranks, which would use the received data to fill in their local part of the full matrix.
Once the broadcast is completed, the next reader would read its assigned file until all files were read and broadcast.
This algorithm proved to be a
bottleneck when assembling large matrices on massively parallel clusters like Blue Waters.

Hence, for this work we extended the reader code to be split in two phases ``reading from disk'' (I/O) and ``completing the matrix'' (communication).
Doing so avoids the need to serialize during the I/O phase. We designed the algorithm to support any desired number of MPI ranks reading concurrently,
as long as the file system can sustain the concurrent reads.
Once the Hermitian matrix is read, the upper half is filled by sending data from those ranks that read the corresponding transposed part.
This completely avoids serialization while reading and introduces only a small amount of serialization while sending data between ranks due to our use of blocking MPI calls.
With this new approach
the total time spent to read and complete the matrix is almost identical to the raw I/O time on the ``reader'' rank reading the largest
file,
indicating that MPI communication is not a significant part of the time budget.
Finally, when using ChASE to diagonalize the parallel matrix, the matrix needs to be converted from being distributed purely along rows (``stripped'') to being distributed in blocks (``blocked'').
For the current usage scenario there is sufficient memory available to store both the striped and the blocked copy of the matrix, permitting the use
of a simple out-of-place algorithm using blocking MPI one-to-one communication calls to redistribute the matrix data among the MPI ranks.
Overall, these changes resulted in a significant speedup
by reducing time spent to read and build the matrix (see Table~\ref{tab:I/O} for detailed results).

\subsection{Solving the BSE eigenvalue problem on massively parallel architectures}

The BSE Hamiltonian matrix is both dense (most of its entry are non-zero) and large with a size $N$ up to $10^6$.
When dealing with dense Hermitian matrices, it is customary to use a so-called ``direct'' eigensolver, which computes solutions by directly reducing the form of the matrix to tri-diagonal form.
Afterwards an iterative solver is used to further transform the matrix to diagonal, to recover its real eigenvalues.

When only a small number $m \ll N$ of low-lying eigenvalues is sought, it is customary to resort to iterative eigensolvers, which are typically used in the case of sparse matrices.
The main reason for this choice resides in the fact that the number of floating point operations needed to compute the solution is $\mathcal{O}(N^2m)$, which can be much smaller than the $\mathcal{O}(N^3)$ operations required by a direct eigensolver.
The break-even point between these two classes of algorithms depends on the number of iterations needed by the iterative algorithm to declare
all desired eigenpairs converged.

The eigenvectors of the BSE Hamiltonian represent physical low-lying eigenmodes and provide information on the single-particle excitations that contribute to a given excitonic state.
Hence, we are interested in the full eigenpairs not just the low-lying eigenvalues.
In this case, it is advisable to use an iterative eigensolver based on a subspace projection.
The main rationale is that such an algorithm deals with the entire set
of desired eigenvectors instead of converging each one in a sequential
fashion, as it is typical for Krylov space methods \cite[Ch. 9]{10.5555/1355328}.
The advantage of subspace methods is that the search space for the eigenvectors can be treated as one contiguous block and further refined at each iteration of the eigensolver.
This is the main philosophy behind the version of the Conjugate
Gradient (CG) method \cite{Geradin, Fried} modified by Kalkreuter and
Simma (KSCG) in their work \cite{kalkreuter1996accelerated}.

The advantage of CG with respect to the classic Lanczos algorithm is in its ability to return eigenvalues with controlled numerical errors and correct multiplicities.
Kalkreuter and Simma improved the CG algorithm by alternating CG minimization with direct minimization of the subspace spanned by the approximate eigenvectors.
The KSCG solver yields a computational cost in terms of FLOP count that scales as $\mathcal{O}(N^2mC_0)$, where $C_0$ is the total number of CG cycles to converge all desired $m$ eigenpairs.
In KSCG the number of needed CG cycles grows with larger eigenvalues and is determined dynamically at runtime.
Therefore it is not possible to have an \emph{a priori} estimate of it.
Nonetheless the total FLOP count is greatly reduced from the $\mathcal{O}(N^3)$ needed for the exact diagonalization of the matrix.

Originally conceived for applications in Quantum Chromodynamics, the KSCG eigensolver parallel implementation is based on geometrical data
decomposition where vectors are equally partitioned and stored on distinct processing nodes.
This choice was based on the assumption that the matrix of the eigenproblem is sparse and local, implying communication only between nearest-neighbor nodes for matrix-vector multiplications.

With the adoption of the KSCG algorithm, our BSE code had been applied to successfully study excitonic properties for many material systems, see e.g.\ Refs.\ \cite{schleife2012ab,Kang:2019} and references therein.
Unfortunately, because the BSE Hamiltonian is dense and lacks a local structure, the KSCG eigensolver ends up having a much larger inter-node
communication load both in terms of message size and number of MPI collectives calls.
We also note that each CG cycle deals with a vector at a time which implies that the arithmetic intensity (the number of operations per byte) of the algorithm is relatively low.
Typically such an algorithm is bound by memory bandwidth and has a performance far from the theoretical peak of the processor.
For these reasons KSCG suffers from limited parallel scalability.
In turn, such a limitation curbs the amount of parallel resources that can be used effectively, and restricts the usage of our BSE code to investigate large physical systems which could be simulated on modern massively parallel architectures with thousands of compute cores.

The challenge of efficient use of parallel computing resources when solving large dense eigenvalue problems is not new.
Several attempts have focused on improving direct methods \cite{imamura2014eigenexa,marek2014elpa,elsi} that are, in part, based on kernels with a high
arithmetic intensity which exploit the multi-core processor performance.
A typical example is the operation of reduction of a dense matrix to
banded form which is based on the {\it fast FLOPs}\footnote{The term
  ``fast FLOPs'' here is the opposite as the term ``slow FLOPs'' already defined in the introduction.} of Basic Linear Algebra Subroutines level 3 (BLAS-3) kernels~\cite{marek2014elpa}.
These kernels are the most optimized in numerical linear algebra
libraries \cite{goto_anatomy_2008, Goto:2008cl, huang_generating_2017} and can achieve up to 95\% of the theoretical peak performance of the processor.
Despite their recent advances all direct solvers are limited by the intrinsic computational complexity of their approach which scales as $\mathcal{O}(N^3)$ and limits their effective parallelization.

Our aim is to keep the advantage of fast FLOPs and to reduce the complexity of the eigensolver.
To our knowledge, the only recent development in the field of numerical linear algebra that can achieve both goals is the Chebyshev Accelerated Subspace iteration Eigensolver (ChASE) \cite{winkelmann2019chase}.
This eigensolver has been initially developed to solve sequences of dense eigenproblems as they arise in Density Functional Theory (DFT) based on
plane waves.
The effectiveness of ChASE stems from a spectral filter based on Chebyshev polynomials.
These polynomials can be computed through a 3-term recurrence relation, which enables the filter to be expressed in terms of BLAS-3 kernels.
In addition, the fraction of computational time spent in the filter is by far larger than any other task within ChASE.
This makes ChASE an extremely efficient solver in terms of node-level performance.

When parallelized over multiple nodes, ChASE takes advantage of its simple algorithmic structure based on subspace iteration and matrix-matrix multiplication to keep the amount of communication to an acceptable level.  
In practice, the BSE matrix is distributed only once across all MPI ranks in equal blocks of data and never re-distributed.
Communication through MPI collective calls involves only the matrix of filtered vectors spanning the subspace.
It has been shown \cite{winkelmann2019chase} that communication overhead is minimized whenever the number of MPI ranks can be arranged in a square Cartesian grid.
Performance is achieved, at the node-level, by using specialized multi-threaded libraries such as MKL or cuBLAS (for computation on GPU cards).
Finally, ChASE allows pre-computing and minimizing the total FLOP count necessary to reach convergence.

Because the ChASE library and the BSE code are written in C++ and Fortran, respectively, the integration and usage of ChASE as the eigensolver necessitated the implementation of a glue code and,
as described in Sec.\ \ref{sec:IOdistr}, the initialization and distribution of the BSE Hamiltonian had to be reworked to be compatible with the data layout of ChASE.
This also reduced the time spent in I/O and redistribution of data.
ChASE 
requires the choice of a variable which sets the size of the search
subspace which is a superset of the subspace spanned by the desired eigenvectors of the BSE matrix.
As we will see in the next section (and in the appendix) such a choice may influence the speedup and convergence.

In the next section we present an exhaustive number of numerical tests executed on two distinct massively parallel clusters.
We illustrate the limitation of the KSCG algorithm, compared to the ChASE eigensolver, and show how ChASE provides much better parallel scalability, which eventually will allow calculations on much more complex materials.


\section{\label{sec:results}Numerical experiments}

In this section we illustrate the results of a series of numerical experiments aimed at addressing the issues discussed above, related to matrix initialization and eigenproblem solutions using massively parallel computing clusters.
We show how the restructuring of the parallel I/O dramatically reduces the time spent reading and setting up the BSE Hamiltonian matrix.
We compare the KSCG and ChASE eigensolvers across a range of eigenproblems.
Finally, we show how the use of ChASE enables our BSE code to further push the accessible physical parameter space, thanks to an enhanced use of parallel resources.
The experiments are divided into three sets, each presenting a unique set up meant to drive our points home.

\paragraph{Strong scaling}
We select a specific BSE eigenvalue problem, such that the size of its
matrix fits into one computing node of a given cluster.
We then initialize the matrix and solve for a small set of the
eigenpairs using both ChASE and KSCG.  This procedure is carried out
for an increasing number of computing nodes to examine the scaling of
the computation 
keeping the data set constant. On a log-log
plot, a positive outcome would result in a linearly decreasing
time-to-solution as a function of computing nodes used.  

\paragraph{Weak scaling}
For these experiments we successively increase the size of the BSE
eigenvalue problem with the number of nodes, keeping the workload per
compute node approximately constant.  This is achieved by changing the
energy cutoff $E_{\rm cut}$ of the BSE Hamiltonian.  We compute the
solutions of these eigenvalue problems with both KSCG and ChASE
eigensolvers and examine the change in computing time.  Good parallel
behavior would result in a roughly constant time to solution as a
function of the number of computing nodes. 

\paragraph{Practical case: Converging the exciton-binding energy}
In the last set of experiments, we showcase a study illustrating how
leveraging a bigger set of computational resources enables us to solve
very large BSE matrices which were previously inaccessible.  A typical
example is provided by organic crystalline naphthalene, for which
converging the exciton-binding energy with respect to the BSE energy
cutoff $E_{\text{cut}}$ requires solving matrices larger than
previously considered.
Thanks to the use of ChASE it was possible to reach the converged regime for the exciton-binding energy of this material.

\paragraph{Resources}
Most of these experiments are performed on the National Science Foundation (NSF) BlueWaters supercomputer, hosted at the National Center for Supercomputing Applications (NCSA) at the University of Illinois, Urbana-Champaign.
On BlueWaters we exclusively used the Cray ``XE'' compute nodes, each equipped with two AMD 6276 Interlagos processors connected via the Gemini interconnect.
Each node has 16 floating point Bulldozer cores and 64 GB of memory.
On BlueWaters our BSE code as well as the KSCG and ChASE eigensolvers are compiled with PGI v13.6.0 and linked against Cray's LibSci v12.1.3, and MPICH v6.1.3.

We repeated a subset of the weak scaling tests on standard CPU and multi-GPU nodes of the JUWELS~\cite{JUWELS} cluster hosted at the J\"ulich Supercomputing Centre (JSC).
Each standard node of JUWELS is equipped with two Dual Intel Xeon Platinum 8168 CPUs (48 cores), an EDR-Infiniband (Connect-X4) interconnect, and 96 GB of memory.
In addition to standard compute nodes, JUWELS is equipped with 56 GPU nodes with two Dual Intel Xeon Gold 6148 CPUs (40 cores), connected via a dual EDR-Infiniband (Connect-X4) and 192 GB of memory each, hosting 4 NVIDIA V100 GPU cards.
Currently, ChASE supports the use of a single GPU device per MPI rank.
GPU experiments on JUWELS use all NVIDIA V100 by running 4 MPI processes per node.
ChASE was compiled with the Intel compiler and Intel MKL version 19.0.3.199, ParaStation MPI version 5.2.2-1, and CUDA version 10.1.105.
In the following, we explain these experiments in detail and analyze the results.

\subsection{\label{sec:strong}Strong scaling}

\begin{table*}[ht]
\centering
\caption{\label{tab:I/O}Strong scaling and speedup of the code with the old I/O routine and KSCG solver, compared to the new parallel I/O implementation and the ChASE solver. Data is collected for runs using between 1 (16 threads) and 64 (1024
  threads) processes.}

\begin{tabular}{c|cccccc}
\toprule
\# MPI ranks& Old I/O&Parallel I/O&Speedup&KSCG&ChASE&Speedup\\ \midrule
1&336.55$\pm$6.10&791.16$\pm$19.69&{\bf 0.43}&5515.65$\pm$4.64&2390.08$\pm$2.59&{\bf 2.31}\\
4&374.96$\pm$9.08&374.26$\pm$7.21&{\bf 1.00}&1554.04$\pm$0.93&617.83$\pm$2.63&{\bf 2.52}\\
9&413.86$\pm$3.11&191.00$\pm$0.85&{\bf 2.17}&830.26$\pm$0.22&270.86$\pm$3.09&{\bf 3.07}\\
16&408.63$\pm$7.45&97.24$\pm$0.76&{\bf 4.20}&592.23$\pm$1.27&161.30$\pm$2.22&{\bf 3.67}\\
25&443.46$\pm$14.78&78.94$\pm$0.76&{\bf 5.62}&471.94$\pm$0.31&109.59$\pm$0.66&{\bf 4.31}\\
36&491.44$\pm$12.28&40.38$\pm$0.34&{\bf 12.17}&427.56$\pm$0.31&82.01$\pm$0.08&{\bf 5.21}\\
49&451.09$\pm$2.46&33.81$\pm$0.71&{\bf 13.34}&383.55$\pm$2.71&67.82$\pm$0.08&{\bf 5.66}\\
64&450.20$\pm$2.10&78.17$\pm$1.32&{\bf 5.76}&359.89$\pm$0.50&55.92$\pm$0.18&{\bf 6.44}\\
\bottomrule
\end{tabular}

\end{table*}

For the strong scaling tests, we selected hafnium oxide, HfO$_2$, with a Brillouin zone sampling of
$6\times 6\times 6$ {\bf k}-points and a BSE energy cutoff of \num{9.1} eV.
This corresponds to a BSE eigenvalue problem with a matrix size of $N=\num{41252}$, and we solve for the $m=100$ lowest eigenpairs of the spectrum.
In order to minimize overhead generated by MPI communicators, we always use a number of MPI ranks which is the square of an integer number.
This is due to the fact that most collective communications internal to ChASE are either AllReduce or Broadcasts which benefit from a square grid of processes;
even better gain is achieved when the square is a power of two.
We measured CPU time to completion within the BSE code for the tasks
related to matrix initialization and eigenproblem solution, using
between 1 and 64 processes \footnote{We use equivalently and
  interchangeably the terms process and MPI rank.}.
For each node of BlueWaters (BW) we run the computation always with a
number of threads equal to the total number of 16 floating point cores per node.

\begin{figure}[h!t]
\centering
\includegraphics[width=0.45\textwidth]{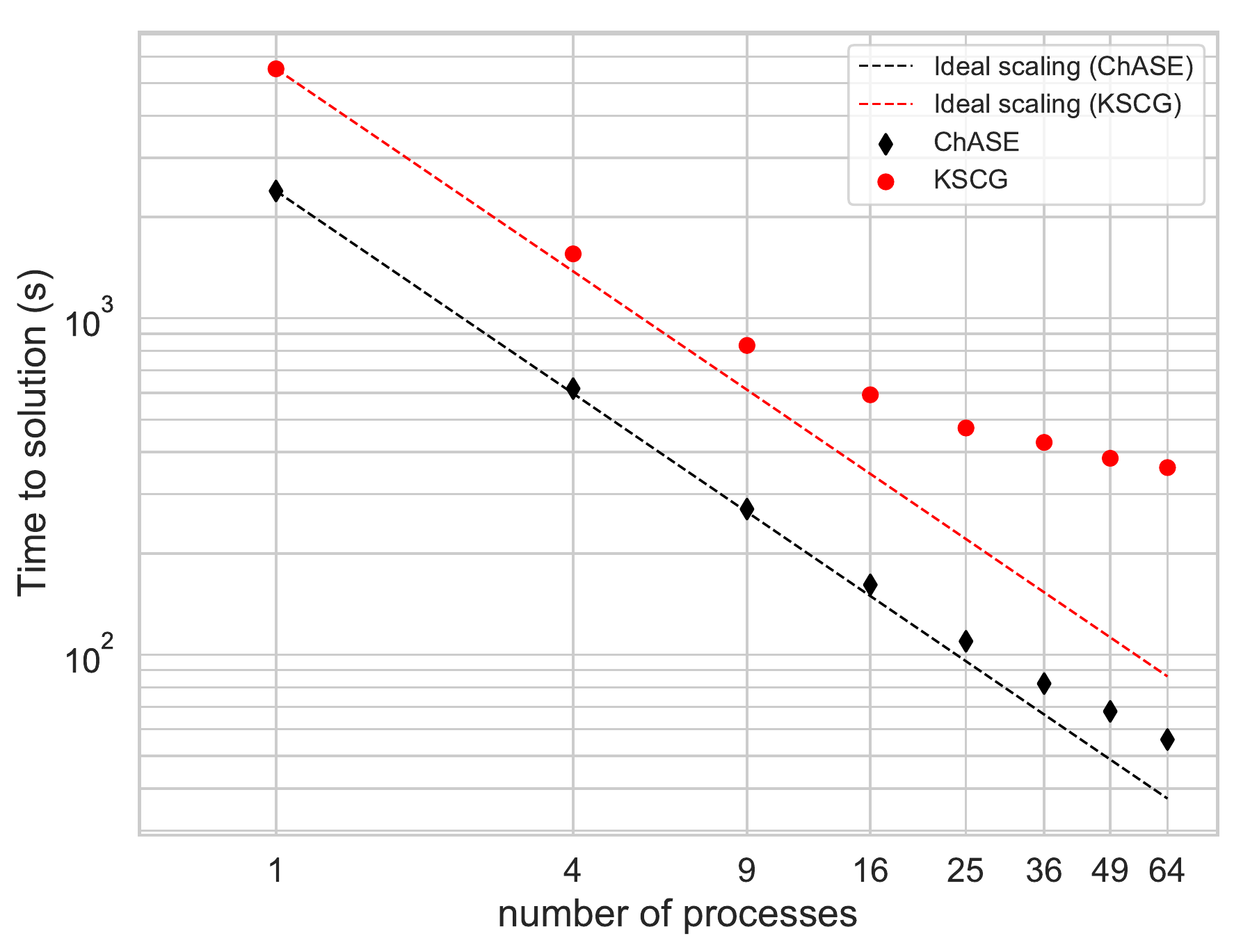}
\caption{\label{fig:time_solver}
  Strong scaling test.
  For a test with a specific number of MPI ranks, the computing time
  is averaged over 8 identical calculations on exactly the same
  computing nodes to account for statistical fluctuations in the
  floating point operations and communications among nodes. 
}
\end{figure}

We first show the benefit of adopting the new reading routine:
Tab.\ \ref{tab:I/O} shows the improvement that the new parallel I/O delivers when compared to the original routine.
In the original routine, only one MPI rank would read at a time, so while different ranks would read different files, only one rank would read at any given time.
In practice, the code was no faster than having a single MPI rank read all files and broadcast the contents.

The new parallel routine scales well with increasing number of MPI ranks:
For 1\,--\,4 MPI ranks, the CPU time of the new I/O is longer or comparable to the sequential routine, due to the additional ``blocked'' re-distribution process.
While KSCG can use the old ``stripped distribution'', this redistribution is needed when invoking ChASE.
Beyond 4 MPI ranks, there is a stark improvement for the parallel routine and the peak speed-up of over 13 occurs for 49 MPI ranks for this example.
The new algorithm is designed so that each MPI rank reads concurrently the part of the matrix that it holds, resulting in the lower half of the matrix being initialized.
Subsequently the upper half is initialized by sending MPI one-to-one messages between the MPI ranks holding an initialized part of the matrix and the to-be-initialized transposed part.
This approach avoids serialization during the I/O phase and only introduces a small amount of serialization among the ranks during the communication phase, reducing the time spent to read the complete matrix to barely more than the time required to read in the segment owned by any of the MPI ranks.



For the strong scaling experiment, and for both the KSCG and ChASE eigensolvers, we set the value ${\sf nex}=30$;
this choice implies that the search space contains 30 additional eigenvalues on top of the wanted part of the spectra (${\sf nev}=100$).
This feature of subspace iteration eigensolvers is necessary in order to avoid that the convergence of the largest desired eigenvectors becomes extremely slow.
In order to demonstrate the scaling of the eigensolver, we show in Fig.\ \ref{fig:time_solver} the time-to-solution
as a function of the number of MPI ranks used.
For each number of MPI ranks, the timings  are averaged over 8 runs.
Already on a single node, ChASE typically solves the eigenproblem in a fraction of the time needed by the KSCG solver.
For the example chosen here, this corresponds to a 2.3x speed up, which grows to 6.4x on 64 nodes (see Tab.\ \ref{tab:I/O}).


The growth in speedup, a consequence of the better scaling of the ChASE eigensolver, can be easily observed by comparing the deviation from the ideal scaling curves (dashed lines in Fig.\ \ref{fig:time_solver});
ChASE remains quite close to ideal scaling up to the largest number of compute nodes we used (\num{64}), but KSCG deviates from ideal scaling already for 16 nodes and is far off for \num{64} nodes.
This behavior is expected, as the KSCG solver experiences an increase
in the communication overhead relative to the time spent in carrying
out actual computations.
Thanks to the use of BLAS-3 kernels, ChASE makes better use of the CPU resources.
Furthermore, it does not experience an early change of the ratio
between computation and communication, until the amount of data per
MPI rank starts becoming too small for the BLAS-3 kernels to take full
advantage of their arithmetic intensity.

\begin{figure}[h!t]
  \centering
  \includegraphics[width=0.45\textwidth]{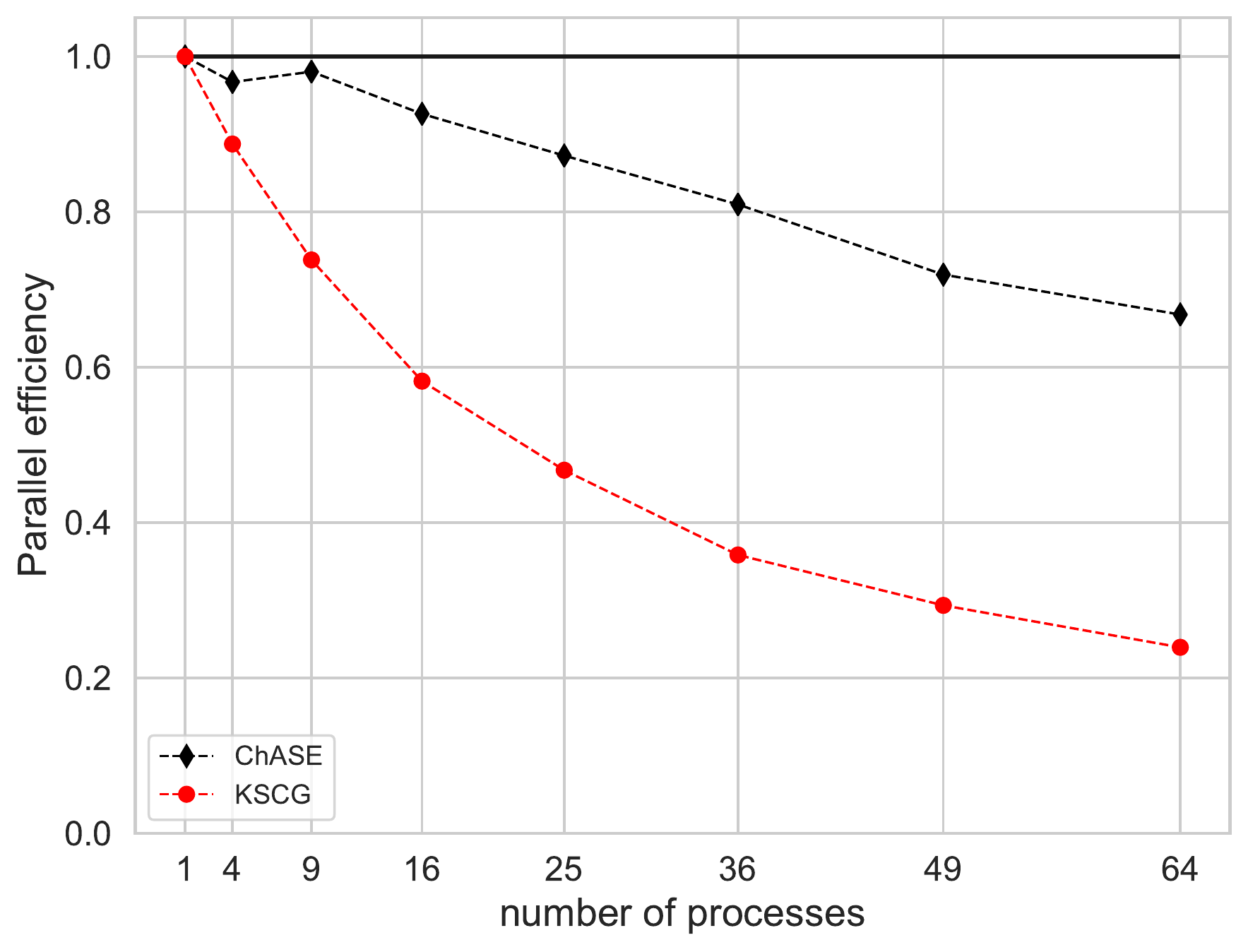}
  \caption{\label{fig:PE}The parallel efficiency of the ChASE solver
    (black) and the CG solver (red) with respect to one compute node
    (16 cores).}
\end{figure}

This behavior is directly quantified by the parallel efficiency $\eta$ of both solvers, which measures the loss in efficiency of a computational task as the number of processing units $p$ increases 
\begin{equation}
\eta=\frac{t_{\text{ref}}*p_{\text{ref}}}{t*p}.
\end{equation}
In our setup, the reference time $t_{\rm ref}$ and processing units $p_{\rm ref}$ refer to a single-node simulation with 16 cores.
The plot of $\eta$ versus the number of processes is shown in Fig.\ \ref{fig:PE}.
It can be seen that the parallel efficiency of ChASE remains above
90\,\% up to 16 nodes, while that of the KSCG solver drops rather rapidly.
For 64 ranks, the efficiency of the KSCG solver drops to 0.24, while that of the ChASE solver is still 0.67, well above the 50\,\% mark.
  
The speed up and parallel efficiency prove that the ChASE eigensolver significantly accelerates the process of obtaining solutions for practically relevant excitonic eigenvalue problems, and offers a better parallel behavior when compared to the KSCG solver.
As mentioned in Sec.\ \ref{sec:compute}, the reason for the substantial speed up of ChASE comes from its use of highly optimized BLAS-3 computational kernels which extract as much performance as possible at the node level.
The better scaling of ChASE resides in a lower communication overhead threshold which makes it more competitive.

\subsection{\label{sec:weak}Weak scaling}

Weak scaling measures the behavior of a computational task when its workload is maintained constant while the computing resources are increased.
In order to realize this set up we solve increasingly large
eigenproblems, commensurate to the growth in number of compute nodes.

In this section, we illustrate the weak scaling of both eigensolvers using the Indium oxide (In$_2$O$_3$) system, that was previously studied in \cite{schleife2018optical}.
The $\mathbf{k}$-point sampling is fixed at $5\times 5\times 5$ and the BSE matrix size is changed through the BSE energy cutoff $E_{\rm cut}$, which determines the number of non-interacting electron-hole pair states involved (see Sec.~\ref{sec:physics}).
Precisely tuning the energy cutoff allows us to adjust the matrix size
from $N\sim\num{38500}$ to $N\sim\num{500000}$, while increasing the
number of compute nodes and keeping the bulk of workload per node
approximately constant.
The matrix sizes corresponding to the different $E_{\text{cut}}$ and the node counts are shown in Tab.~\ref{tab:ecutoff}.

Unfortunately, one cannot exactly predict the total workload to reach convergence for an iterative eigensolver, but it is reasonable to assume that its order of magnitude is $\mathcal{O}(N^2)$, as shown in Sec.~\ref{sec:compute}.
Based on this assumption, the workload per node is kept within $\pm$0.5\% of that for the simulation on a single node.
With this set of parameters, the utilization of the memory for each single node across different matrix sizes is kept at $\sim$80\,\% and $\sim$50\,\% in the BW cluster and JUWELS cluster, respectively.

The weak scaling tests executed on both clusters have slightly different goals:
On BW we tested larger matrices with sizes up to half a million and compared the scaling between the KSCG and the ChASE solvers.
On JUWELS we scanned a lower number of nodes and focused on just the ChASE eigensolver, executing it on CPU nodes and GPU equipped nodes.
$E_{\rm cut}$, the matrix size, and the number of MPI ranks are reported in Table \ref{tab:ecutoff} and we kept all other eigensolver parameters the same, i.e., {\sf nev}=100 and {\sf nex}=25.
On BW we kept the number of MPI ranks the same as the number of nodes and set the number of OMP threads per rank equal to the total number of computing cores in each node.
Because the JUWELS cluster has more cores per node and four GPU devices on each GPU node, we used 4 MPI ranks per node with 12 and 10 OMP threads per rank on the CPU nodes and GPU nodes, respectively.

\begin{table*}[ht]
  \centering
  \caption{\label{tab:ecutoff}
    $E_{\rm cut}$ and matrix sizes used for the weak scaling tests on the Blue Waters and JUWELS clusters as a function of the computing nodes.
  }

\begin{tabular}{c|cc|ccc|cc}
\toprule
nodes&\multicolumn{2}{c|}{Blue Waters}&\multicolumn{3}{c|}{JUWELS}&Energy cutoff&Matrix size \\
    &MPI&OMP&MPI&OMP(CPU)&OMP(GPU)&(eV)&($N$) \\ \midrule 
1&1&16&4&48&40&6.45&\num{38537}\\
4&4&64&16&192&160&7.45&\num{76887}\\
9&9&144&36&432&360&8.31&\num{115459}\\
16&16&256&64&768&640&9.11&\num{154023}\\
25&25&400&100&1200&1000&9.87&\num{192788}\\
36&36&576&144&1728&1440&10.54&\num{231011}\\
49&49&784&196&2352&1960&11.15&\num{269645}\\
64&64&1024& & & &11.73&\num{307865}\\
81&81&1296& & & &12.30&\num{346915}\\
100&100&1600& & & &12.82&\num{385183}\\
121&121&1936& & & &13.32&\num{423607}\\
144&144&2304& & & &13.83&\num{462469}\\
169&169&2704& & & &14.30&\num{500649}\\ \bottomrule
    \end{tabular}
\end{table*}


\begin{figure}[h]
    \centering
    \includegraphics[width=0.45\textwidth]{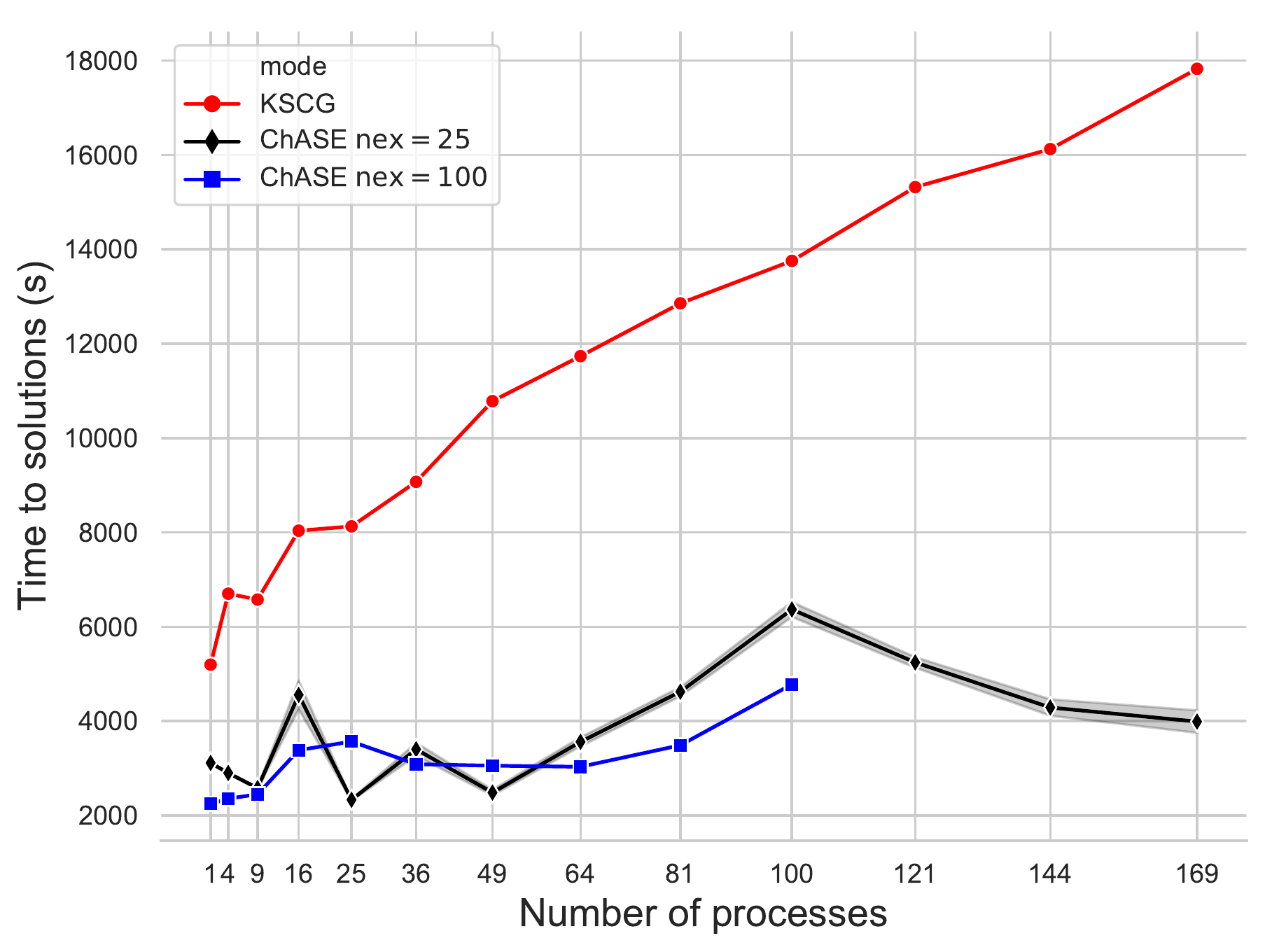}
    \caption{\label{fig:BW_weakscaling}
      Weak scaling test.
      Averages over 5 runs are performed for ChASE with $n_{\text{ex}}=25$ and the error bars are plotted.
    }
\end{figure}

Figure \ref{fig:BW_weakscaling} illustrates weak scaling of ChASE and KSCG via the total time needed to solve for the first 100 eigenpairs as a function of the number of computing nodes used on BW.
In order to determine variability, we executed this test on BW 5 times
for each of the distinct matrix sizes and solvers and averaged over the
measured completion times.  While for ChASE this procedure was
repeated for all node configurations, for the KSCG solver this average is
only performed up to 81 nodes due to the high computational cost and
the small standard deviation. 
Since the KSCG error bars were no larger than 0.2\% across all
calculations and are not visible in the plot, only ChASE standard
deviation is shown. We also repeated the numerical tests for ChASE
with a different value of {\sf nex}=100 and discuss motivation and
results in Sec.~\ref{sec:conv}.

Figure \ref{fig:BW_weakscaling} shows that for KSCG the increase in
time is almost monotonic and for ChASE it follows an oscillating
behavior as the number of processes increases.  For the KSGC solver,
the maximum time of execution (\num{17825} secs on 169 nodes) is about
3.4 times longer than the minimum (\num{5197} secs on a single node).
Correspondingly, the longest time of execution (\num{6522} secs on 100
nodes) for ChASE is only 2.67 times longer than the shortest time
(\num{2443} secs on 25 nodes), indicating a better weak scaling of the
ChASE solver. While the smallest and largest time-to-solution for both
solvers occur for different node counts, it is important to notice
that the ratio between the minima and maxima of the two solvers
increases from 2.1 to 2.7. This increasing trend is analogous to the
increase in speed-up between the two solvers in the strong scaling
case, indicating a better parallel efficiency for the ChASE
eigensolver.  This trend is even more evident for node counts above
100 nodes: the speed up of ChASE with respect to KSCG keeps increasing
consistently, reaching 4.3 at 169 nodes.


The timing for ChASE varies especially in the range between 16 and 64
nodes, corresponding to matrix sizes between 10$^5$ and
3$\times 10^5$. These oscillations are in part algorithmic and in part
due to variation in the communication overhead. The algorithmic causes
of oscillations originate in the iterative nature of the ChASE
solver. Because iterative numerical algorithms are not deterministic,
they do not have a constant work per byte of data. In other words, one
cannot predict a-priori the number of operations necessary to reach a
solution. In the specific case of ChASE, what varies is 1) the number
of subspace iterations, and 2) the different counts of matrix-vector
operations required to converge across eigenproblems of different size
(see supplementary material for more details). Both these factors
directly determine the effective workload associated with an
eigenproblem solution by ChASE.

The communication overhead is mostly
due to two MPI collectives used in the ChASE implementation: AllReduce
and Broadcast. We illustrate their effect in more detail later in this
section when discussing experiments executed on JUWELS.  In
conclusion, the numerical behavior of KSCG and ChASE is influenced by
similar factors having different effects on both solvers. Time spent in
inter-node communication tends to increase timings for both solvers as
the number of computing nodes grows. The less regular behavior of
ChASE is additionally determined by the changes in the effective
workload executed by the solver.

\begin{figure}[h]
  \centering
  \includegraphics[width=0.4\textwidth]{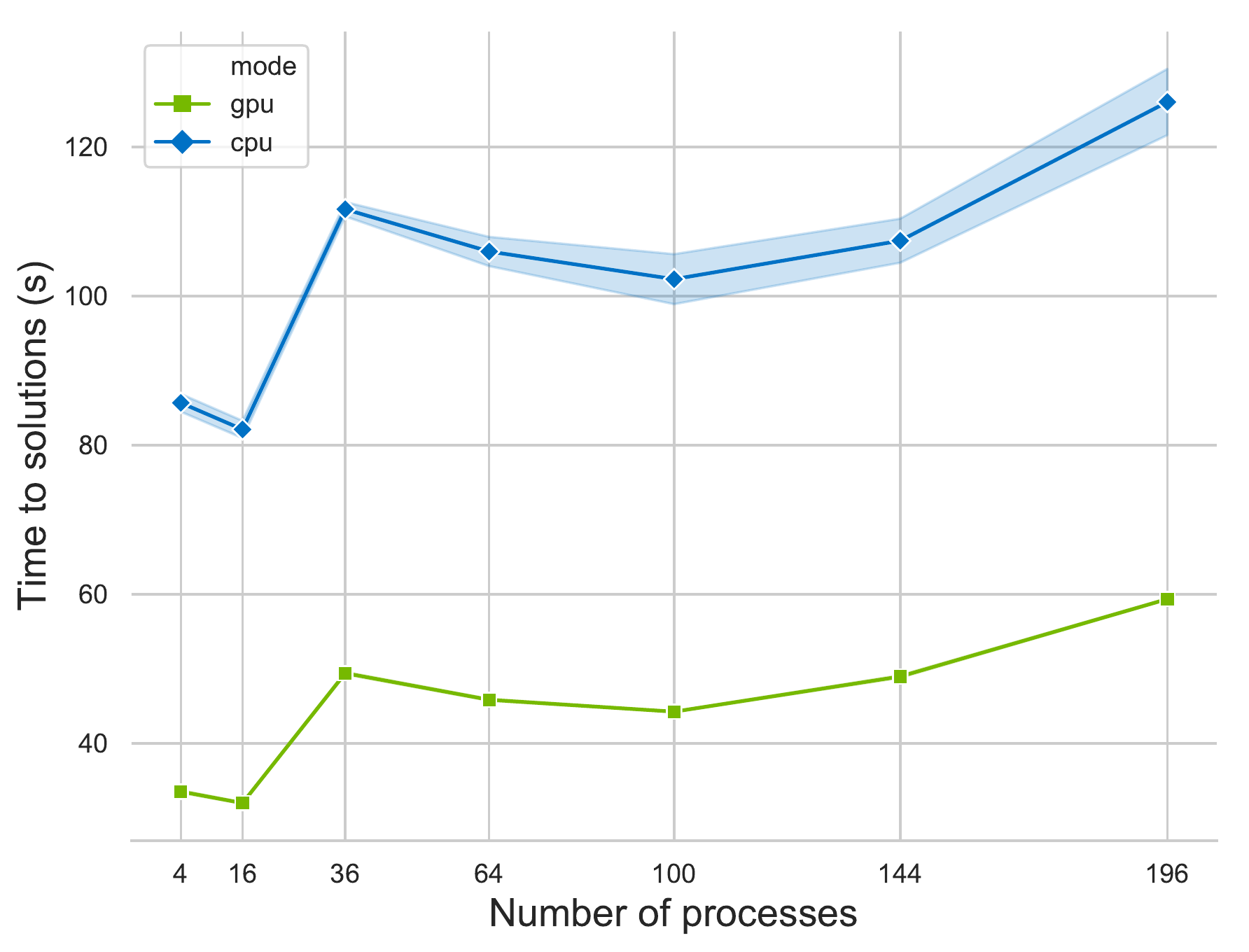}
  \caption{\label{fig:JW-weakscaling}
    Plot of weak scaling results for the ChASE solver on
    JUWELS for both CPU (blue) and GPU (green) nodes. Statistics are
    over 15 runs. The plot shows average and 95\,\% confidence
    interval based on standard deviation in shade of the same color.}
\end{figure}

The results of the tests executed on the JUWELS cluster~\cite{JUWELS} are shown in
Fig.~\ref{fig:JW-weakscaling}.  The parameters of ChASE execution were
the same as for Blue Waters, namely {\sf nev}=100 and {\sf nex}=25.
To match the number of GPU devices hosted by JUWELS GPU nodes, we use
a different configuration of the computing resources by assigning four
MPI ranks to each node. In order to improve the statistical analysis
of our results we executed 15 repetitions for any given matrix size on
both CPU and GPU nodes. For each point in the plot we show the average
and its 95\,\% confidence interval (CI) based on the standard
deviation.  The results are shown as a set of points, blue indicating
time-to- solution for the CPU nodes and green for the GPU nodes,
encased by a shaded area indicating the CI.

While the standard deviation for CPU nodes is much larger than for GPU
nodes, the time-to-solution on CPU nodes and GPU nodes also shows
noticeable similarities of the pattern of variation as a function of
the number of processes.  In particular, a big jump can be observed in
going from 16 to 36 processes.  This is a typical case where there are
two factors both adding to the computing time: a drastic increase in
the total matrix-vector operations executed by ChASE and a substantial
increase in latency of the many calls to MPI {\tt AllReduce}.  We have
benchmarked such call (as well as the {\tt Ibcast} call) and observed
a jump from 15.6 ms to 21.9 ms in average latency.  Likewise, ChASE
performs \num{8160} and \num{11360} matrix-vector multiplications
using 16 and 36 MPI ranks, respectively (see complete set of data in
the supplementary material).  This latter effect is not always
positive but it may also favor a decrease in run time as it can be
deduced observing that the number of matrix-vector operation decreases
from \num{11360} to \num{9340} in going from 36 to 100 MPI ranks, respectively.
On top of these two effects, the average latency of MPI Bcast keeps
increasing, contributing with a constant ratio to the overall increase
in time-to-solution.

Overall, it is clear that, whenever possible, an execution over
multiple GPU devices should be preferred, since in the worst-case
scenario it halves the run time with respect to the use of only CPU
nodes.  On the other hand, the peak performance of the four NVIDIA
V100 GPUs is significantly larger than two times that of the CPU
nodes, which implies that the use of the GPUs is not as efficient as
on the multi-cores. This effect is expected since cuBLAS is less
efficient than the Multi-threaded MKL BLAS in dealing with repeated
multiplications between a square and a tall and skinny matrix.  In
addition, there is some unavoidable overhead due to the transfer of
the filtered vectors from the main to the local GPU memory and vice versa.

\subsection{\label{sec:conv}Search space and ChASE convergence}

As shown in Ref.\ \cite{winkelmann2019chase}, the convergence of ChASE can be significantly affected by the choice of the number of additional eigenpairs {\sf nex}, which together with the wanted part of the spectra form the full search space.
Specifically, in ChASE the convergence of a given eigenpair $(\lambda_a, x_a)$ is deterministic and is related to the distance between $\lambda_a$ and the center of the interval $[\lambda_{\sf nev+nex},\lambda_{\textsf{max}}]$ divided by its half-width.
In our weak-scaling test the minimum eigenvalue $\lambda_1$ changes
very little, while the maximum eigenvalue $\lambda_{\textsf{max}}$
gets significantly larger as the size of the matrices increases.  In
turn, such a change leads to a worsening of the convergence ratios of
the desired eigenpairs (see example for the correlation between the
convergence of the solver and the matrix size in the supplemental
information).  The convergence can be improved by increasing the {\sf
  nex} value such that $\lambda_{\sf nev+nex}$ is increased
accordingly.  However, as more extra eigenvalues are added, the
convergence of the iterative solver is improved at the expense of
increasing the workload due to a larger number of the vectors that
iterate and need to be filtered.  In an attempt to obtain a less oscillatory
curve for the weak scaling for ChASE on BW, we now analyze how the behavior
of the solver can be influenced by the {\sf nex} value as the matrix
size increases.

To illustrate the effect of the {\sf nex} value, we measured the
time-to-solution for the same weak scaling problem set, but with
${\sf nex}=100$.
In Fig.\ \ref{fig:BW_weakscaling} we compare these results up to
100 nodes and observe that increasing {\sf nex} leads to a more flat curve
of the time-to-solution for ChASE.  Between 9 and 64 nodes it can
be clearly seen that the run time oscillates less, compared to
${\sf nex}=25$. In addition, the three largest matrices benefit
substantially from choosing a larger {\sf nex} showing a consistent decrease in timing.
Intermediate size matrices do not benefit in the same way.
As mentioned above, this is attributed to the trade-off between two
competing factors; the performance gain obtained from a better
convergence rate and the increased in size of the search space which causes a growth in workload.

To prove this point, we picked the matrix of size $N=\num{231011}$ and
measured the run time as a function of {\sf nex}. Indeed we observe an
optimized value of {\sf nex} around 40\,--\,50 (see details in the
supplemental information). A potential way to improve the choice of
{\sf nex} across different matrices is to perform such a test once, and
then predict the {\sf nex} value accordingly across different matrix
sizes.  Predicting a scaling factor in a systematic way, however, is
not straightforward, and can be affected significantly by the
distribution of eigenvalues in the overall spectra.  Thus, it can
depend significantly on the number of desired eigenpairs.

\subsection{\label{sec:phys_problem}A practical case: converging exciton-binding energy in naphthalene}

\begin{figure}[ht]
\centering
\includegraphics[width=0.4\textwidth]{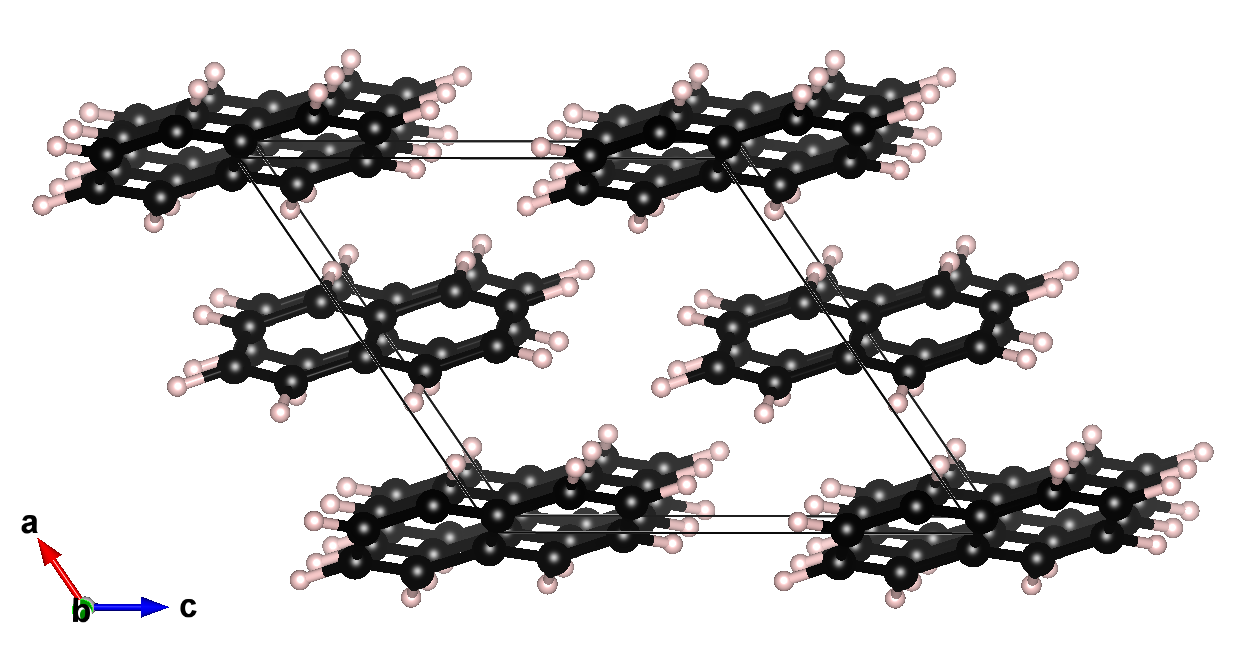}
\caption{\label{fig:crystalstr}Primitive unit cell of naphthalene crystal with black spheres being carbon atoms and white spheres being hydrogen atoms. $a, b$, and $c$ labels the crystal axis. }
\end{figure}

In this section, we illustrate how the adoption of ChASE enables us to
address physical problems which were inaccessible before, due to their high computational cost.
In particular, we show that using ChASE enables to solve for extremely large eigenvalue problems that occur when accurately converging the exciton binding energy of a crystal with respect to the BSE energy cutoff.
Here we use the example of a naphthalene organic crystal (see Fig.\ \ref{fig:crystalstr} for the crystal structure of the material):
In this system, the exciton-binding energy converges slowly with respect to the BSE energy cutoff
and requires matrices with ranks up to $\sim\num{500000}$.
This is only possible thanks to the excellent weak scaling behavior of the ChASE eigensolver.

\begin{table*}[ht]
\centering
\caption{\label{tab:naph_size}
  BSE matrices for the convergence test of the exciton binding energy (in eV) for a naphthalene organic crystal with respect to the BSE energy cutoff (in eV).
  As the size increases (with increasing energy cutoff) we use more processes with the ChASE solver.
  Compute timings (in s) and the binding energies of the dark and optically active excitonic state are reported (see text).}
\begin{tabular}{l|ccccc}
\hline
1/$E_{\text{cut}}$ &size        & \# processes & time (s)& E$_{\text{b, dark}}$ & E$_{\text{b, active}}$ \\ \hline
0.1667 &7172&16&10.672&1.18224&0.93402\\
0.125 &22264&16&69.087&1.2053&0.97823\\
0.1         & 49888        & 36   & 176.678  & 1.23005    & 1.03827 \\
0.0833 & 92220        & 36   & 626.752  & 1.23496    & 1.06192 \\
0.0714 & 149808       & 36   & 2829.956 & 1.23945    & 1.09635 \\
0.0625      & 217292       & 36   & 3316.959 & 1.24473    & 1.13995 \\
0.0556 & 296648       & 64   & 3882.796 & 1.24878    & 1.1908  \\
0.05       & 381364       & 121  & 4443.877 & 1.25251    & 1.23184 \\
0.0455 & 472444       & 169  & 4256.253 & 1.25671    & 1.26287 \\ \hline
\end{tabular}
\end{table*}

Naphthalene is an organic crystal with 20 carbon atoms and 16 hydrogen atoms per primitive unit cell.
We treat carbon\,$2s$ and carbon\,$2p$ as valence electrons, leading to a total of 96 valence electrons.
Although this unit cell is not as large as that for In$_2$O$_3$, there are noteworthy difficulties in obtaining exciton-binding energies:
The lowest three bound excitonic states are dark, which means that their optical oscillator strength, representing the strength of the dipole transition, is low.
In practice, the exciton binding energy in this material is typically defined as the difference between the lowest peak of the imaginary part of the dielectric function
without and with considering excitonic effects~\cite{hummer2005}.
The lowest peak with excitonic effects included corresponds to the fourth lowest eigenvalue of the BSE matrix.
Thus, we study the convergence of both optically dark and active states with respect to BSE energy cutoff in this work.

To this end, we compute the lowest eigenvalue (dark excitonic state) and the fourth-lowest eigenvalue (optically active excitonic state of the first peak for light polarization along the $y$ direction).
We report the size of the eigenproblem, the run time for ChASE,
and the calculated exciton binding energies in Tab.\ \ref{tab:naph_size}.
It can be seen from the table that as the energy cutoff increases (decreasing $1/E_{\textbf{cut}}$), the matrix size increases significantly and reaches $\sim\num{500000}$ for the largest energy cutoff we simulated.
Using the KSCG solver, it was not possible to tackle matrices of this size, and the largest matrix that we can reach was on the order of $\sim\num{200000}$, corresponding to $E_{\text{cut}}=16$ eV.

\begin{figure}[h]
  \centering
  \includegraphics[width=0.45\textwidth]{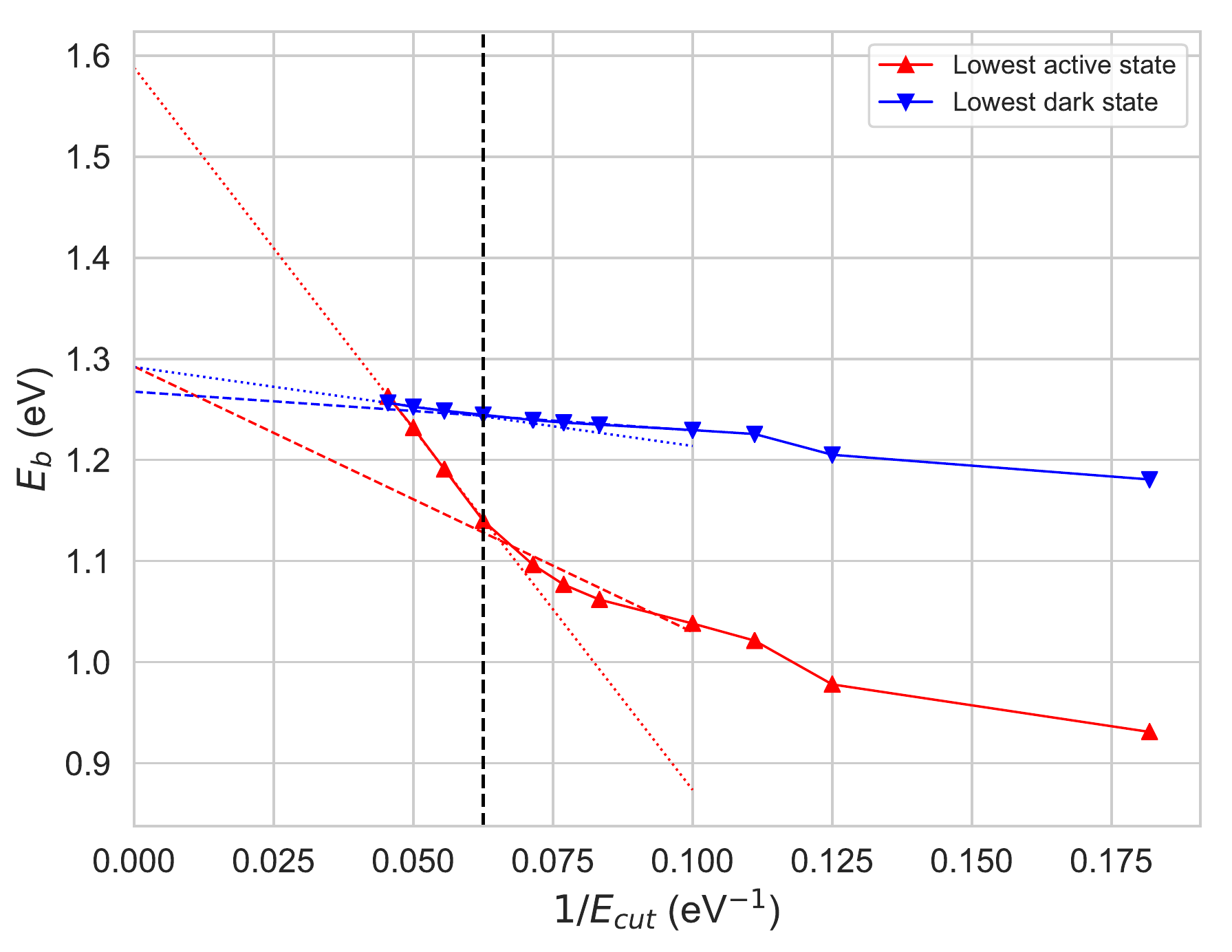}
  \caption{\label{fig:naph_eb}
    Convergence of the exciton-binding energy of the lowest dark state (blue) and first optically active state (red) with respect to the BSE energy cutoff $E_{\text{cut}}$.
    Extrapolation of the curve to zero results in the exciton binding energy and is shown using data points from $E_{\text{cut}}$=10\,--\,16 eV (dashed line, red and blue) and $E_{\text{cut}}$=18\,--\,22 eV (dotted line, red and blue). The black dashed vertical line represents the limit of KSCG solver. 
  }
\end{figure}

The convergence test in Fig.\ \ref{fig:naph_eb} shows that by pushing the matrix size beyond what was possible using the KSCG solver, we find a great improvement of the convergence of exciton-binding energy for this material.
As mentioned in Sec.\ \ref{sec:physics}, the converged exciton-binding energy is obtained through linear extrapolation of the corresponding eigenvalues of matrices computed using the largest values of $E_{\text{cut}}$.
In particular, we compare extrapolations of $E_{\rm b}$ obtained using data points from $E_{\text{cut}}$=10\,--\,16 eV (achievable using the KSCG solver), and from $E_{\text{cut}}$=18\,--\,22 eV (only achievable using the ChASE library).

This figure shows that for both the dark and the optically active state, the linear regime starts emerging only after matrix sizes $N\sim\num{200000}$, 
corresponding to $E_{\text{cut}}=16$ eV.
Significant differences can be seen between the extrapolation results using the two sets of data.
For the dark state, using only data points in the range of $E_{\text{cut}}=$10\,--\,16 eV results in an extrapolated value of $E_{\text{b}}=1.267$ eV, compared to a value of $E_{\text{b}}=1.292$ eV obtained using data points in the range $E_{\text{cut}}$=18\,--\,22 eV.
The difference is much larger for the optically active state, with the extrapolation of $E_{\text{b}}=1.293$ eV ($E_{\text{cut}}=$10\,--\,16 eV) vs.\ $E_{\text{b}}=1.588$ eV ($E_{\text{cut}}=$18\,--\,22 eV).
In other words, using eigenvalues
exclusively computed with the KSCG algorithm underestimates the value of the extrapolated exciton binding energy by 2\,\% and 18.6\,\%, for the lowest dark and lowest active state, respectively.
The underestimation of the exciton-binding energy for the lowest
active state is not only large, but also plays a more important role
from a practical point of view.
The lowest optically active state corresponds to the onset of measurable optical absorption and impacts the use of a material, e.g.\ for optoelectronic applications, more than the lowest dark state.
Many similar materials exhibit optically dark lowest excitonic states and the lowest optically active states higher in energy \cite{takagi2013,congreve2013external}.

In summary, using naphthalene as an example, we show that increasing the size of the BSE Hamiltonian is critical to obtain an accurate extrapolation of the exciton-binding energy.
Integrating the ChASE solver into our code allows us to solve eigenproblems with matrix sizes two to five times larger than what was possible before.
Adoption of the ChASE solver, thus, enables
more accurate predictions of the exciton-binding energy and opens up opportunities to study the optical properties of materials more accurately.
This can benefit many applications such as organic solar cells, for which knowledge of the exciton binding energy is amongst the most important aspects to be uncovered~\cite{congreve2013external}.


\section{\label{sec:conclusion}Summary and Conclusions}

We described and analyzed the modernization of a legacy theoretical spectroscopy code.
This code allows for \emph{in silico} predictions of optoelectronic properties of materials by solving the Bethe-Salpeter equation for the optical polarization function.
Here we identified two main bottlenecks, namely the initialization of the underlying excitonic Hamiltonian matrix from disk and the solution of the corresponding eigenvalue problem for a small fraction of the lowest part of the spectrum.
We addressed each bottleneck individually:
First we re-factor and parallelize the I/O routines that are responsible for reading the matrix elements from files and distributing them among the compute nodes.
Second, we integrate the scientific software ChASE, a new parallel library specialized in solving for partial spectra of large and dense Hermitian eigenproblems.

Our results show that the new parallel I/O routine scales well with the number of compute nodes and in our test we find a speed-up of up to 13.34 compared to the legacy implementation.
This was determined when reading a matrix of size \num{41252} on 49
compute nodes, i.e., a total of about 842 rows per MPI rank, below which increasing inter-node communication reduces the speedup.

To characterize the implementation of ChASE for finding the lowest eigenvalues, we carry out three sets of performance measurements:
Strong scaling tests, weak scaling tests, and tests on the convergence of the excitonic binding energy of organic crystalline naphthalene.
The last set of tests illustrates for a material of practical importance that the modernized code can access exciton physics at a scale that was previously unattainable.
Using these results we demonstrate that the integration of the new
ChASE library into the BSE code achieves two main objectives: First
it enhances both the parallelism and the scaling behavior of the
scientific software, allowing it to be used on the current and future
parallel clusters. Second, it extends the ability of the software
to tackle scientific questions which could not be answered so far with the necessary accuracy.

This is a success story of how to modernize a code by changing both its algorithmic structure and the middleware embedded in it.
Because the significantly higher computational cost of the previous
implementation imposed accuracy limitations, it became a necessity to
reduce such costs by addressing the existing bottlenecks.
The end result is a rejuvenated BSE code that will allow its users to study novel problems at scales inaccessible before.
Overall, the modernization of this scientific software has extended its lifetime especially in light of current computing cluster moving towards the exascale target.

This work describes the results of a long term project initiated under the Joint Laboratory for Extreme Scale Computing (JLESC), an international consortia of Supercomputing Centers and Research institutions.
The interdisciplinary nature of the challenge at the core of this project brought together researchers from diverse fields as condensed matter physics, numerical linear algebra, and high-performance computing across two continents.
The results obtained are a demonstration how effective is a multi-disciplinary team in addressing a computational challenge that requires a variety of expertise.

\section*{Acknowledgments}

This material is based upon work supported by the National Science Foundation under Grant No.\ DMR-1555153.
This research is part of the Blue Waters sustained-petascale computing project, which is supported by the National Science Foundation (awards OCI-0725070 and ACI-1238993) and the state of Illinois.
Blue Waters is a joint effort of the University of Illinois at Urbana-Champaign and its National Center for Supercomputing Applications.
The authors acknowledge the computing time granted through
J\"ulich Supercomputing Centre on the supercomputer JUWELS at
Forschungszentrum J\"ulich. Financial support from the Deutsche
Forschungsgemeinschaft (DFG) through grant GSC 111 is also gratefully acknowledged. 
This research is partially supported by the NCSA-Inria-ANL-BSC-JSC-Riken-UTK Joint-Laboratory for Extreme Scale Computing (JLESC, \url{https://jlesc.github.io/}), which enabled joint workshop attendance for scientific discussions and a visit of E.\ D.\ N.\ and J.\ W.\ at NCSA in Urbana-Champaign.

\bibliographystyle{elsarticle-num}
\bibliography{ChASE.bib}

\clearpage

\setcounter{section}{0}
\setcounter{equation}{0}
\setcounter{figure}{0}
\setcounter{table}{0}
\setcounter{page}{1}
\makeatletter
\renewcommand{\thesection}{S\arabic{section}}
\renewcommand{\theequation}{S\arabic{equation}}
\renewcommand{\thetable}{S\arabic{table}}
\renewcommand{\thefigure}{S\arabic{figure}}
\renewcommand{\thepage}{S\arabic{page}}
\renewcommand{\bibnumfmt}[1]{[S#1]}
\renewcommand{\citenumfont}[1]{S#1}

\onecolumn

\section{Supplemental materials}
\label{sec:supplemental}

In this section we report on additional data relative to the numerical
tests described and discussed in Sec.~\ref{sec:results}. Most of the
data in presented in table formats with few additional comments on the
data format and its significance.

\subsection{Strong scaling tests}

Table~\ref{tab:ave} and Table~\ref{tab:old_time} report the full
timings recorded for the ChASE and the CG solver,
respectively. The timing is broken into multiple main parts:

\begin{itemize}
\item {\bf Pre-reading:} Before reading any matrices or actual data,
  these indicate the task of reading the input parameters,
  initialization, or allocations, these are collected in one single timer.
\item {\bf Read optics:} After the initialization, the optical
  transition matrix elements, as well as the DFT energies are read
  from the ground state calculation of the material system.
\item {\bf Reading and building:} This label indicates the task of
  proper reading and initializing the BSE matrix. In the paper we have
  used the term I/O for this task. This is the first
  bottleneck for which the code was modernized and where we observe
  substantial improvement with respect to the old code.
\item {\bf CG/ChASE:} This label corresponds to the time spent in
  obtaining the lowest set of eigenpairs of the BSE Hamiltonian either
  using the Conjugate gradient solver or
  ChASE library. This is the second and most substantial bottleneck of
  the BSE code and where we observe a marked 
  difference with respect to the old code.
\item {\bf Total:} This label indicates the sum of all timings above. 
\end{itemize}

\begin{table*}[ht]
\centering
\caption{\label{tab:ave}Strong scalability test of the code with new reading routine and ChASE solver. For all different runs, 16 threads are used per MPI rank and the data is averaged over 8 runs on the same node(s)
}
\begin{tabular}{c|ccccc}
\toprule
\# MPI ranks&Pre-reading&Read optics&Reading/Building&ChASE&Total\\ \midrule
1&26.26$\pm$0.04&26.29$\pm$0.01&791.16$\pm$19.69&2390.08$\pm$2.59&3235.15$\pm$19.43\\
4&29.91$\pm$0.13&30.03$\pm$0.01&374.26$\pm$7.21&617.83$\pm$2.63&1052.15$\pm$7.96\\
9&29.91$\pm$0.02&30.01$\pm$0.01&191.00$\pm$0.85&270.86$\pm$3.09&521.81$\pm$3.35\\
16&30.01$\pm$0.09&30.01$\pm$0.01&97.24$\pm$0.76&161.30$\pm$2.22&318.54$\pm$2.12\\
25&30.08$\pm$0.11&30.04$\pm$0.01&78.94$\pm$0.76&109.59$\pm$0.66&248.65$\pm$1.04\\
36&30.16$\pm$0.04&30.02$\pm$0.01&40.38$\pm$0.34&82.01$\pm$0.08&182.58$\pm$0.38\\
49&30.00$\pm$0.05&30.02$\pm$0.01&33.81$\pm$0.71&67.82$\pm$0.08&161.65$\pm$0.75\\
64&30.03$\pm$0.08&30.02$\pm$0.01&78.17$\pm$1.32&55.92$\pm$0.18&194.14$\pm$1.38\\
\bottomrule
\end{tabular}
\end{table*}

The average value and standard error of the timings are obtained
through averaging over 8 runs on the same computing node(s). The
columns corresponding to the solvers and the I/O are repeated in
Table~\ref{tab:I/O}. For the sake of completeness we have reported
here also timings for Pre-reading and Read Optics. The timings in
these two columns refer to an intrinsically sequential part of the
code and do not scale. They also account for a fraction of the two
bottlenecks and that is why we do not have included or addressed them
in the paper.   

\begin{table*}[ht]
\centering
\caption{\label{tab:old_time}Strong scalability for the old reading and building routine and the KSCG solver. 16 number of threads are used per MPI rank and the data is averaged over 8 runs on the same node(s). }
\begin{tabular}{c|ccccc}
\toprule
\# MPI ranks&Pre-reading&Read optics&Reading/Building&KSCG&Total\\ \midrule
1&25.89$\pm$0.02&25.81$\pm$0.03&336.55$\pm$6.10&5515.65$\pm$4.64&5903.90$\pm$8.59\\
4&29.69$\pm$0.14&29.55$\pm$0.03&374.96$\pm$9.08&1554.04$\pm$0.93&1988.24$\pm$8.75\\
9&29.92$\pm$0.44&30.12$\pm$0.08&413.86$\pm$3.11&830.26$\pm$0.22&1304.17$\pm$3.48\\
16&29.68$\pm$0.07&29.63$\pm$0.06&408.63$\pm$7.45&592.23$\pm$1.27&1060.17$\pm$7.75\\
25&29.75$\pm$0.09&29.59$\pm$0.06&443.46$\pm$14.78&471.94$\pm$0.31&974.74$\pm$14.74\\
36&29.72$\pm$0.07&30.02$\pm$0.06&491.44$\pm$12.28&427.56$\pm$0.31&978.30$\pm$12.30\\
49&29.66$\pm$0.05&29.84$\pm$0.01&451.09$\pm$2.46&383.55$\pm$2.708&894.15$\pm$3.53\\
64&29.71$\pm$0.05&29.63$\pm$0.11&450.20$\pm$2.10&359.89$\pm$0.50&869.43$\pm$2.20\\
\bottomrule
\end{tabular}
\end{table*}

\subsection{\label{sup:weakscaling}Weak scaling tests}

In this part, we show the data for the CPU-times for the weak scaling
tests. For almost all simulations on BW, the average and standard
deviation of the time to solution are computed over 5 runs on the
same node(s). Only the simulations execute on 100 or larger computing
nodes using the old code and solver have been run only once. Due to
the long duration, these latter executions were very costly in terms
of computing time. Moreover, the variation in timings also for
smaller number of computing nodes was so negligible that repeating the
measurement would have brought no benefit to the statistics of the
result.  All executions are performed with one MPI rank and 16 ranks
per computing node.


\begin{table*}[ht]
\centering
  \caption{\label{sup:weak_results}Results of the weak scaling tests
    on Blue Waters for both the old and new I/O and both solvers,
    ChASE and KSCG. All simulations were executed with one MPI rank
    per node}
\begin{tabular}{@{}c|cccccc@{}}
\toprule
MPI ranks & Matrix size  & ChASE time (s)   & STDEV (s) & KSCG time (s) &STDEV (s)      & Speed up \\ \midrule
1           & 38537  & 3188.35 & 10.89  & 5197.46 &1.28 & 1.63      \\
4           & 76887  & 2986.44 & 14.61  & 6701.22 &2.06 & 2.24      \\
9          & 115459 & 2672.35 & 21.04  & 6576.31 &1.99 & 2.46      \\
16          & 154023 & 4657.37 & 303.26 & 8036.14 &1.37 & 1.73      \\
25          & 192788 & 2443.02 & 17.20  & 8128.18 &9.97 & 3.33      \\
36          & 231011 & 3522.34 & 138.67 & 9072.10 &1.75 & 2.58      \\
49          & 269645 & 2615.52 & 61.03  & 10781.91&19.69 & 4.12      \\
64         & 307865 & 3699.07 & 96.77  & 11736.11&2.99 & 3.17      \\
81         & 346915 & 4771.18 & 97.73  & 12855.16&6.50 & 2.69      \\
100         & 385183 & 6522.20 & 158.07 & 13755.39&- & 2.11      \\
121         & 423607 & 5408.97 & 111.74 & 15318.24&- & 2.83      \\
144         & 462469 & 4462.82 & 174.62 & 16123.97&- & 3.61      \\
169         & 500649 & 4172.64 & 237.95 & 17825.44&- & 4.27      \\ \bottomrule
\end{tabular}
\end{table*}

In Table~\ref{tab:weak_juwels_CPU} and \ref{tab:weak_juwels_GPU} we
report average timings and corresponding standard deviation for the
procedures internal to the ChASE eigensolver over 15 repetitions on
the JUWELS cluster. The
labels on the columns indicates respectively:
\begin{itemize}
\item {\bf Lanczos:} This a modified Lanczos algorithm to compute the
  approximate spectral density which is used to estimate the value of
  $\lambda_1$, $\lambda_{\sf nev+nex}$ and $\lambda_N$. This procedure
  is executed only once, the first time the solver is invoked.
\item {\bf Filter:} The Chebyschev filter is the computational core of
  the solver and the most intensive in terms of FLOPs. This is also
  the procedure that is most efficient since it is practically a
  repeated call to the {\tt HEMM} subroutine of the BLAS library. This
  routines and all the remaining below are executed at each internal
  {\it while} loop of ChASE. Each repetition of the loop roughly corresponds
  to an iteration of the subspace projection.
  \item {\bf QR:} This column corresponds to a QR decomposition of the
    filtered vectors outputted by the Filter. It is executed redundantly on each node.
  \item {\bf RR:} The Rayleigh-Ritz step corresponds to the projection
    unto the subspace spanned by the filtered vectors $Q$ outputted by
    the QR decomposition. It also includes the solution of the reduced
    eigenproblem through a standard solver from the LAPACK library and
    a back-transformation of the approximate eigenvectors. Both the
    projection and the back-transformation are executed by repeated
    invocation of {\tt GEMM} subroutines.    
  \item {\bf Resid:} This label indicates the computation of the
    eigenpairs residual and the deflation and locking of converged
    vectors. Also in this case most of the computation is carried on
    using {\tt GEMM}.
\end{itemize}

\begin{table*}[ht]
\centering
\caption{\label{tab:weak_juwels_CPU}Timings of the weak scaling tests
  on JUWELS for the CPU nodes. All values are averaged over $15$ runs.}
\begin{tabular}{@{}c|cccccc@{}}
\toprule
\# MPI (p)   & ChASE time (s)    & Lanczos (s)      & Filter (s)       & QR (s)          & RR (s)          & Resid (s) \\
\midrule
4            & $ 85.68 \pm 1.19$ & $18.83 \pm 0.42$ & $58.07 \pm 0.99$ & $0.90 \pm 0.01$ & $2.90 \pm 0.05$ & $2.88 \pm 0.05$       \\
16           & $ 82.12 \pm 1.16$ & $19.17 \pm 0.84$ & $53.31 \pm 0.75$ & $1.80 \pm 0.04$ & $3.07 \pm 0.06$ & $3.03 \pm 0.06$       \\
36           & $111.66 \pm 0.96$ & $19.75 \pm 0.86$ & $76.28 \pm 0.79$ & $4.37 \pm 0.08$ & $4.43 \pm 0.08$ & $4.39 \pm 0.13$       \\
64           & $105.99 \pm 1.96$ & $21.34 \pm 1.61$ & $67.71 \pm 0.64$ & $5.79 \pm 0.13$ & $4.58 \pm 0.09$ & $4.47 \pm 0.19$       \\
100          & $102.28 \pm 3.36$ & $23.00 \pm 2.96$ & $62.09 \pm 0.35$ & $6.40 \pm 0.15$ & $4.66 \pm 0.20$ & $4.36 \pm 0.13$       \\
144          & $107.43 \pm 2.96$ & $24.05 \pm 2.94$ & $63.67 \pm 0.67$ & $8.10 \pm 0.24$ & $5.04 \pm 0.20$ & $4.83 \pm 0.17$       \\
196          & $126.05 \pm 4.45$ & $28.69 \pm 4.13$ & $71.37 \pm 0.43$ & $11.91 \pm 0.24$ & $6.21 \pm 0.29$ & $5.78 \pm 0.15$       \\\bottomrule
\end{tabular}
\end{table*}

All the calls to the BLAS subroutines are executed on the computing
node either by calls to the corresponding multi-threaded routines of
the Intel MKL (Table~\ref{tab:weak_juwels_CPU}) or the Nvidia CuBLAS
library (Table~\ref{tab:weak_juwels_GPU}). The only exception is the
QR decomposition which is always executed on the CPU and not on the
GPU cards. One can observe the effect of such implementation in the
increasing values of the timings in the QR column as the number
of computing nodes gets larger. This is a limitation of the current
algorithm in the ChASE library. Future version of ChASE will feature a
distributed QR decomposition and an automatic mechanism which would
switch from a redundant node-level execution to a full distributed
one.

\begin{table*}[ht]
\centering
\caption{\label{tab:weak_juwels_GPU}Timings of the weak scaling tests
  on JUWELS for the GPU nodes. All values are averaged over $15$ runs.}
\begin{tabular}{@{}c|cccccc@{}}
\toprule
\# MPI (p)   & ChASE time (s)   & Lanczos (s)     & Filter (s)       & QR (s)          & RR (s)          & Resid (s)  \\
\midrule
4            & $33.54 \pm 0.19$ & $3.82 \pm 0.04$ & $17.37 \pm 0.16$ & $1.04 \pm 0.02$ & $0.63 \pm 0.04$ & $0.54 \pm 0.01$       \\
16           & $31.99 \pm 0.10$ & $3.90 \pm 0.07$ & $15.87 \pm 0.06$ & $2.08 \pm 0.03$ & $0.92 \pm 0.02$ & $0.79 \pm 0.01$       \\
36           & $49.42 \pm 0.18$ & $4.00 \pm 0.05$ & $25.05 \pm 0.12$ & $5.45 \pm 0.07$ & $1.61 \pm 0.05$ & $1.47 \pm 0.05$       \\
64           & $45.85 \pm 0.13$ & $4.19 \pm 0.04$ & $20.88 \pm 0.11$ & $6.93 \pm 0.07$ & $2.00 \pm 0.04$ & $1.69 \pm 0.04$       \\
100          & $44.25 \pm 0.08$ & $4.21 \pm 0.05$ & $19.48 \pm 0.08$ & $7.83 \pm 0.04$ & $2.27 \pm 0.04$ & $2.00 \pm 0.09$       \\
144          & $48.97 \pm 0.18$ & $4.36 \pm 0.07$ & $21.18 \pm 0.09$ & $9.82 \pm 0.05$ & $2.77 \pm 0.05$ & $2.37 \pm 0.05$       \\
196          & $59.36 \pm 0.09$ & $4.45 \pm 0.07$ & $24.26 \pm 0.08$ & $13.85 \pm 0.04$ & $3.55 \pm 0.08$ & $3.16 \pm 0.11$       \\ \bottomrule
\end{tabular}
\end{table*}

In Table~\ref{tab:weak_juwels_iter} we report the details of the runs
of both CPU and GPU nodes of the JUWELS cluster including the number
of subspace iterations and mat-vec multiplications performed. These
mat-vec are a breakdown of all the single multiplications of the BSE
matrix with the filtered vectors. Despite being counted as single
mat-vec for the purpose of having a measure of the complexity of the
execution, all the operations are carried on by multiplying the
BSE matrix with a block of vectors. This strategy is at the base of
the multiple invocation of the BLAS level 3 routines. The ChASE
algorithm, despite being an iterative algorithm, is quite
deterministic. Once a grid of computing nodes has being assigned, the
solver performs the same number of iterations and mat-vec
multiplications to reach convergence independently if it is executed
only on CPU cores or on a hybrid combination of CPU and GPU cores. In other words the
complexity of the algorithm does not change across computing
platforms. 

\begin{table*}[ht]
\centering
\caption{\label{tab:weak_juwels_iter}Nodes configuration and number of
  iteration and mat-vec multiplications for the weak scaling tests on
  JUWELS for the CPU and GPU nodes.}
\begin{tabular}{@{}c|cc|cc|ccc@{}}
\toprule
\# MPI  & \multicolumn{4}{c|}{Node level configuration}  & Matrix & Number of
                                                                       & mat-vec \\
ranks  & \multicolumn{2}{c|}{CPU} & \multicolumn{2}{c|}{GPU} &   size & Iterations & multiplications\\
           & \# OMP & \# GPU & \# OMP & \# GPU                   &                  &              & \\ \midrule
4          & 12     & 0      & 10     & 4                        & 38537            & 6            &   8960       \\
16         & 12     & 0      & 10     & 4                        & 76887            & 5            &   8160       \\
36         & 12     & 0      & 10     & 4                        & 115459           & 7            &  11360       \\
64         & 12     & 0      & 10     & 4                        & 154023           & 6            &  10020       \\
100        & 12     & 0      & 10     & 4                        & 192788           & 5            &   9340       \\
144        & 12     & 0      & 10     & 4                        & 231011           & 5            &   9540       \\ 
196        & 12     & 0      & 10     & 4                        & 269645           & 6            &  10620       \\ \bottomrule
\end{tabular}
\end{table*}

In Table~\ref{tab:MPI_bench_JUWELS} we report benchmark timings of the
two MPI calls that are used within the Filter procedure using the OSU
Micro-Benchmarks\cite{liu2004microbenchmark}. Since this procedure
takes a large percentage of the computing time these benchmarks give a
reasonable measure of the communication overhead experienced by the
ChASE library.
As pointed out in Sec.~\ref{sec:weak}, a number of factors
contributes to the fluctuation of the total time to solution of the
ChASE solver. It is the combination of the variation of the total
number of mat-vec, subspace iterations and duration of the latency of
the calls to {\tt MPI\_Allreduce} and {\tt MPI\_Ibcast} that
contribute to the trend depicted in Fig.~\ref{fig:BW_weakscaling}. For
instance, the latency time for the {\tt Ibcast} increases steadily as
more MPI ranks are used, contributing positively to the total
time. The same cannot be said for the latency of the {\tt Allreduce}
which benefits from a number of MPI ranks equal to a powers of two. In
particular this combination of factors explains the jump in timings
from 16 to 36 nodes due to all factors contributing positively, and
especially the number of mat-vec performed.

\begin{table}[ht]
\centering
\caption{\label{tab:MPI_bench_JUWELS}Results of the Benchmark of
  MPI\_Allreduce and MPI\_Ibcast on the JUWELS cluster obtained using
  the OSU Micro-Benchmarks.}
\begin{tabular}{@{}cc|ccc@{}}
\toprule
\#nodes & MPI & Message size   & Avg. Latency ($\mu$s) & Avg. Latency ($\mu$s)\\
        & Ranks &              & {\tt MPI\_Allreduce}  & {\tt MPI\_Ibcast} \\
\midrule  
 1      &   4        & $16 \text{MB}$ & $12540.32$ & $ 9482.64$              \\
 4      &  16        & $16 \text{MB}$ & $15661.46$ & $17102.91$              \\
 9      &  36        & $16 \text{MB}$ & $21918.39$ & $20391.34$              \\
16      &  64        & $16 \text{MB}$ & $17021.43$ & $22392.87$              \\
25      & 100        & $16 \text{MB}$ & $23168.14$ & $29695.72$              \\
36      & 144        & $16 \text{MB}$ & $21936.58$ & $34585.03$              \\
49      & 196        & $16 \text{MB}$ & $23259.20$ & $41008.10$              \\ \bottomrule
\end{tabular}
\end{table}



\subsection{\label{sup:conv}Dependence of the convergence rate of
  ChASE on {\sf nex}}

In this subsection, we illustrate the importance of choosing an
${\sf nex}$ value that maximizes the convergence rate of ChASE without
excessively increasing the complexity of the total execution. As an
illustrative example we report in Table~\ref{tab:nex25} and
\ref{tab:nex100} the value of some important parameters computed at
the last step of the subspace iteration using both a value
${\sf nex}=25$ and ${\sf nex}=100$ for the weak scaling test on BW.
\begin{itemize}
\item $\lambda_{\sf nev}$ is the value of the last desired
eigenvalue computed by ChASE. In both tables ${\sf nev}=100$.
\item $\alpha$ is an estimate of $\lambda_{\sf nev+nex}$
\item $\beta$ is an estimate of $\lambda_{N}$. This is
computed by the Lanczos procedure and does not change for the
remainder of the ChASE execution. 
\item $c$ and $e$ are, respectively, the center and the half-width of
  the interval filtered out. They are computed from the values of
  $\alpha$ and $\beta$.
\item $\rho_{\sf nev}$ is the rate of convergence of the
  latest desired eigenpair. This is computed according to the
  following formula.
\begin{equation}
   |\rho_{\sf nev}|=\max_{\pm}\Bigg\{\bigg|\frac{\gamma-c}{e}\pm\sqrt{(\frac{\gamma-c}{e})^2-1}\bigg|\Bigg\}
\end{equation}
  Because of the design of the Chebyshev
  filter, such convergence rate is the worst among all sought after
  eigenpairs. We report the inverse of such number which represent the
  dampening factor: the closest to 1, the slower the eigenpair
  $(\lambda_{\sf nev}, x_{\sf nev})$ would converge to a residual below
  the required tolerance threshold.
\end{itemize}
We use values computed during the last iteration since they are the
closest to the corresponding unknown true value of the parameter.

First, the dampening factor $\frac{1}{\rho_{\sf nev}}$ for
${\sf nex}=100$ is in general smaller than that for ${\sf
  nex}=25$. This is expected and corresponds to better convergence of
the iterations process as can be seen by the systematic lower number
of iteration to convergence for ${\sf nex}=100$. Within the same
${\sf nex}$, the rate of convergence shows an increasing trend as the
number of computing nodes increases. Such behavior can be mainly
attributed to the significant increase in $\beta$ as the matrix sizes
increases. Unlikely the $\beta$ values, the
$\alpha\approx\tilde{\lambda}_{\sf nev+nex}$ values show fluctuations
as the size of the matrix increases, the changes in both values leads
to the variation in the dampening factor which, in turn, influences
the time-to-solution for the ChASE solver, as already pointed out is
Sec.~\ref{sec:weak}. The observations above should be considered
qualitative, as they only focus on the largest desired eigenvalue, and
on the last iteration. In practice, the actual convergence of
the entire subspace of eigenvectors is a bit more complicated, and it
is influenced by the affective convergence rate of all the desired
eigenvalues at each iteration.  



\begin{table*}[ht]
\centering
\caption{\label{tab:nex25}Detailed data for the convergence ratio for
  the largest desired eigenvalue $\lambda_{\sf nev}$ of the last step
  of iteration with ${\sf nex}=25$.} 
\begin{tabular}{@{}cccccccc@{}}
\toprule
\# nodes&$\lambda_{\sf nev}$&$\alpha$&$\beta$&$c$&$e$&$1/\rho_{\sf nev}$&\# iterations\\ \midrule
1   & 3.64939 & 3.75894 & 10.08705 & 6.92299  & 3.16406  & 0.76921 & 6  \\ 
4   & 3.64930 & 3.73778 & 13.43655 & 8.58716  & 4.84939  & 0.82635 & 5  \\
9   & 3.64924 & 3.76372 & 14.38214 & 9.07293  & 5.30921  & 0.81278 & 7  \\
16  & 3.64915 & 3.73978 & 16.38626 & 10.06302 & 6.32324  & 0.84442 & 7  \\
25  & 3.64914 & 3.80541 & 18.13631 & 10.97086 & 7.16545  & 0.81183 & 5  \\
36  & 3.64910 & 3.73653 & 19.15953 & 11.44803 & 7.71150  & 0.86033 & 6  \\
49  & 3.64909 & 3.73887 & 20.79428 & 12.26658 & 8.52770  & 0.86504 & 5  \\
64  & 3.64905 & 3.73834 & 21.19633 & 12.46734 & 8.72900  & 0.86683 & 6  \\
81  & 3.64904 & 3.73833 & 22.62650 & 13.18241 & 9.44409  & 0.87162 & 10 \\
100 & 3.64902 & 3.74087 & 22.91544 & 13.32816 & 9.58728  & 0.87083 & 10 \\
121 & 3.64903 & 3.76850 & 24.78667 & 14.27758 & 10.50909 & 0.86016 & 8  \\
144 & 3.64903 & 3.81612 & 26.08816 & 14.95214 & 11.13602 & 0.84113 & 7  \\
169 & 3.64899 & 3.81917 & 26.75923 & 15.28920 & 11.47003 & 0.84194 & 7  \\ \bottomrule
\end{tabular}
\end{table*}

\begin{table*}[ht]
\centering
\caption{\label{tab:nex100}Detailed data for the convergence ratio for
  the largest desired eigenvalue $\lambda_{\sf nev}$ of the last step of
  iteration with ${\sf nex}=100$.  
}
\begin{tabular}{@{}cccccccc@{}}
\toprule
\# nodes&$\lambda_{\sf nev}$&$\alpha$&$\beta$&$c$&$e$&$1/\rho_{\sf nev}$&\# iterations\\ \midrule
1   & 3.64939 & 7.17276 & 10.08705 & 8.62991  & 1.45715 & 0.14956 & 3 \\
4   & 3.64930 & 7.85694 & 13.43655 & 10.64674 & 2.78981 & 0.20797 & 3 \\
9   & 3.64924 & 8.61066 & 14.38214 & 11.49640 & 2.88574 & 0.19055 & 3 \\
16  & 3.64915 & 3.90693 & 16.38626 & 10.14660 & 6.23966 & 0.75091 & 4 \\
25  & 3.64914 & 4.33603 & 18.13631 & 11.23617 & 6.90014 & 0.64238 & 4 \\
36  & 3.64910 & 4.33000 & 19.15953 & 11.74476 & 7.41477 & 0.65355 & 4 \\
49  & 3.64909 & 4.32190 & 20.79428 & 12.55809 & 8.23619 & 0.66932 & 4 \\
64  & 3.64905 & 3.98738 & 21.19633 & 12.59186 & 8.60448 & 0.75615 & 4 \\
81  & 3.64904 & 3.91196 & 22.62650 & 13.26923 & 9.35727 & 0.78938 & 4 \\
100 & 3.64902 & 3.97604 & 22.91544 & 13.44574 & 9.46970 & 0.76947 & 8 \\ \bottomrule
\end{tabular}
\end{table*}

\subsection{\label{sec:optmiz}Optimization of the $nex$ value}

\begin{figure}[h]
    \centering
    \includegraphics[width=0.50\textwidth]{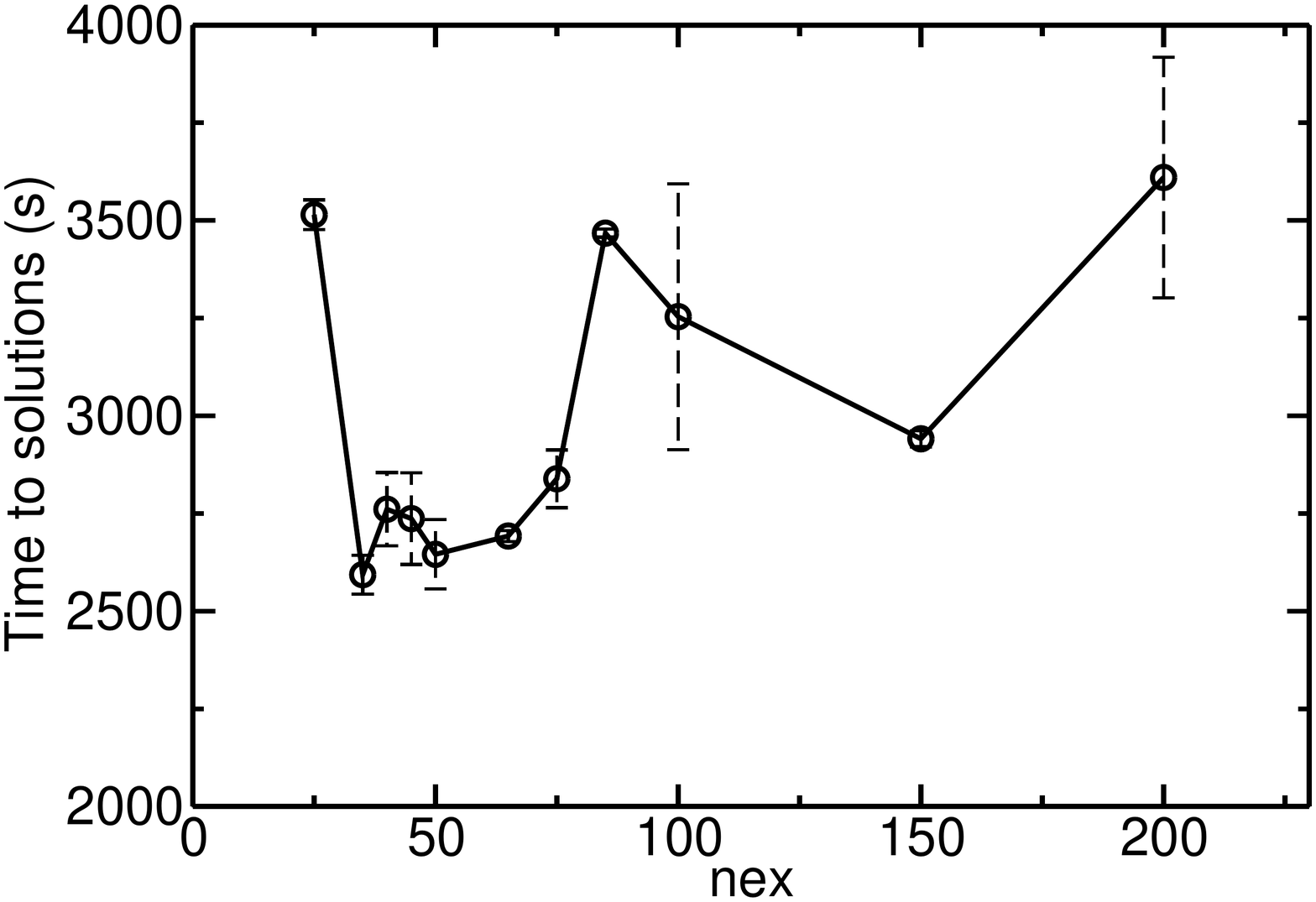}
    \caption{Optimization of the ${\sf nex}$ value for the 36 nodes matrix in the weak scaling test. All data points are averaged over three runs, with the standard deviation plotted in dashed lines. The computational walltime first drops due to better convergence, then increases due to large size of the eigenvectors. 
    }
    \label{fig:optmiz_36}
\end{figure}

In this section, we describe a simple method to optimize the choice of
the ${\sf nex}$ value for the sequence of eigenproblems of varying
size used in the weak scaling test. First, we select a matrix at the
beginning of the sequence for which the value ${\sf nex}=25$ may not
be ideal. This happen to be the eigenproblem with matrix size
$N=\num{231011}$ corresponding to an $E_{\text{cut}}=10.54$
eV. We repeatedly solve this eigenproblem, each time slightly
increasing the ${\sf nex}$ value. We expect to find a value of
${\sf nex}$ for which the computing time of ChASE is minimized. Result
of such a test are shown in Fig.\ref{fig:optmiz_36}. Inspecting the
plot, the interval ${\sf nex}=35-50$ seems to be the one within which
the time-to-solution is likely to be minimal. We artificially set such
a minimal value to be $40$. We define such a value to be the lower end
of a linear regression of ${\sf nex}$ and label it as ${\sf nex_{\rm
    low}}$. Correspondingly we label $E_{\text{cut}}=10.54$ as
$E_{\text{cut-low}}$. In order to avoid consuming too much
computational time, we limit our regression to
$E_{\text{cut}}=12.82$, which we label as $E_{\text{cut-high}}$, and
calculate the corresponding values of ${\sf nex}$ for intermediate
$E_{\text{cut}}$ values using the following formula
\begin{equation}
{\sf nex}={\sf nex}_{\text{low}}+\frac{{\sf nex}_{\text{high}}-{\sf nex}_{\text{low}}}{E_{\text{cut-high}}-E_{\text{cut-low}}}\times(E_{\text{cut}}-E_{\text{cut-low}}).
\end{equation}

\begin{table*}
\centering
\caption{\label{tab:nex_dyn}${\sf nex}$ values computed via linear
  interpolation between ${\sf nex}_{\text{low}}=40$ and ${\sf nex}_{\text{high}}=100, 150$, respectively. 
}
\begin{tabular}{@{}cccc@{}}
\toprule
\# nodes (processors) & $E_{\text{cut}}$ (eV)&${\sf
                                               nex}_{\text{high}}=100$&${\sf
                                                                        nex}_{\text{high}}=150$\\
  \midrule
36 & 10.54 & 40 & 40\\
49 & 11.15 & 50 & 58\\
64 & 11.73 & 59 & 75\\
81 & 12.30 & 68 & 91\\
100 & 12.82 & 76 & 106\\ \bottomrule
\end{tabular}
\end{table*}

\begin{figure}[h]
    \centering
    \includegraphics[width=0.70\textwidth]{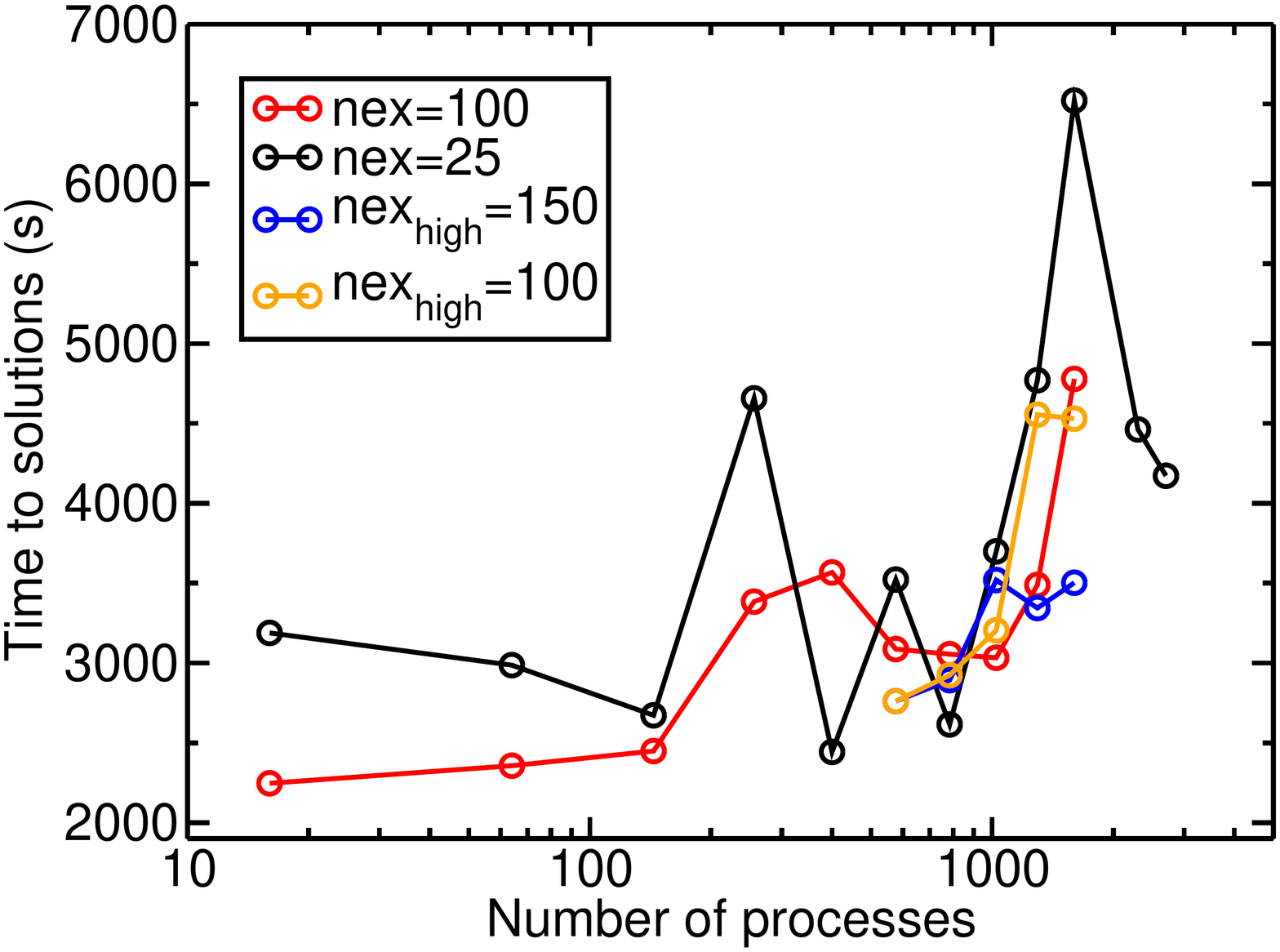}
    \caption{The effect of varying ${\sf nex}$ value across different
      matrix sizes with linear interpolation (orange and blue) and the
      comparison with the weak scaling test with fixed ${\sf nex}$
      (black and red). For both orange and blue curves, ${\sf nex}_{\text{low}}=40$, obtained through the test in Fig.\ref{fig:optmiz_36}. 
    }
    \label{fig:dyn_nex}
\end{figure}

Not knowing a priori the value of ${\sf nex_{\rm high}}$, we choose
two arbitrary values for ${\sf nex}_{\text{high}}=100, 150$, and
report the regressed value of ${\sf nex}$ for intermediate values of
$E_{\text{cut}}$ in Tab.\ref{tab:nex_dyn}. We then tested these values
by solving the corresponding eigenvalue problems with them and
comparing the time-to-solution to the one measured for
${\sf nex}=25,100$ across all problems. The results are plotted in
Fig.\ref{fig:dyn_nex}. We observe that the ${\sf nex}$ values
calculated by regressions present an overall improvement with respect
to the computing time when ${\sf nex}=25$ is used. However, the
same cannot be said when ${\sf nex}=100$ is used across the
board.

While this experiment shows the potential to estimate the correct
value of ${\sf nex}$ for increasing sizes of BSE Hamiltonians, it also
illustrate the non-linear behavior of the
solver as the matrix become larger. 
In general, it is advisable to
increase the ${\sf nex}$ value as the matrix size increases, due to the fact
that the largest eigenvalue $\lambda_{N}$ tends to increase,
leading to the worsening of the convergence rate.  However, linear
scaling of the ${\sf nex}$ value is probably not ideal. The
effectiveness of ${\sf nex}$ is mostly influenced by the distribution
of gaps in the spectrum of the eigenproblem, which is by no means a
linear function. Nonetheless, maintaining the same value of ${\sf
  nex}$ across the board is definitely not advisable and a rule of
thumb based on some sort of simple regression model can save computing
time for the user of ChASE.





\end{document}